\documentclass[preprint, showpacs, reprintnumbers, amsmath, amssymb, superscriptaddress]{revtex4-2}

\usepackage{epsf}
\usepackage{graphicx}
\usepackage{sidecap}

 \usepackage{soul}
 \usepackage{array}
 \usepackage{amsmath}
 \usepackage{amssymb}
 \usepackage{dcolumn}
 \usepackage{epstopdf}
 \usepackage{bm}
 \usepackage{gensymb}
 \usepackage{amsmath}

\usepackage{color}
\usepackage{hyperref}
\usepackage{soul}
\sethlcolor{green}
\hypersetup{
    colorlinks=true,
    citecolor=red,
    linkcolor=red,
    filecolor=blue,   
    urlcolor=blue,
}
 
\def \beq {\begin{equation}}
\def \eeq {\end{equation}}
\begin{document}

\onecolumngrid

\begin{center}
 
  \textbf{\Large Observation of gapless nodal-line states in  $\mathrm{\textbf{NdSbTe}}$}\\[.2cm]
  Sabin~Regmi$^{1}$, Robert~Smith$^{1}$, Anup~Pradhan~Sakhya$^{1}$, Milo~Sprague$^{1}$, Mazharul~Islam~Mondal$^{1}$, Iftakhar~Bin~Elius$^{1}$, Nathan~Valadez$^{1}$ Andrzej~Ptok$^{2}$, Dariusz~Kaczorowski$^{3}$, Madhab~Neupane*$^{1}$\\[.2cm]
 {\itshape
    $^{1}$Department of Physics, University of Central Florida, Orlando, Florida  32816, USA\\
  	$^{2}$Institute of Nuclear Physics, Polish Academy of Sciences, \\W. E. Radzikowskiego 152, PL-31342 Krak\'{o}w, Poland\\
 	$^{3}$Institute of Low Temperature and Structure Research, \\Polish Academy of Sciences, Ok\'{o}lna 2, PL-50-422 Wroc\l{}aw, Poland
  }
\\[.2cm]
$^*$Corresponding author: madhab.neupane@ucf.edu
\\[1cm]
\end{center}
 
\begin{abstract}
{Lanthanide ($Ln$)-based systems in the $\mathrm{ZrSiS}$-type nodal-line semimetals have been subjects of research investigations as grounds for studying the interplay of topology with possible magnetic ordering and electronic correlations that may originate from the presence of $Ln~4f$ electrons. 
In this study, we carried out a thorough study of a $Ln\mathrm{SbTe}$  system - $\mathrm{NdSbTe}$ - by using angle-resolved photoemission spectroscopy along with first-principles calculations and thermodynamic measurements. We experimentally detect the presence of multiple gapless nodal-line states, which is well supported by first-principles calculations. A dispersive and an almost non-dispersive nodal-line exist along the bulk $\mathrm{X}-\mathrm{R}$ direction. Another nodal-line is present well below the Fermi level across the $\mathrm{\Gamma}-\mathrm{M}$ direction, which is formed by bands with high Fermi velocity that seem to be sensitive to light polarization. Our study provides an insight into the electronic structure of an alternative $Ln\mathrm{SbTe}$ material system that will aid towards understanding the connection of $Ln$ elements with topological electronic structure in these systems. } 
\end{abstract}
\maketitle

\begin{center} \textbf{I. INTRODUCTION} \end{center}
Experimental discovery of topological semimetal (TSM) phases has unleashed a flurry of research investigations in semimetallic quantum materials. The family of TSMs has ever since expanded to Dirac semimetals \cite{TSM, Na3Bi, Cd3As2}, Weyl semimetals \cite{TSM, TaAsXu, TaAsDing, Weyl, WeylII}, nodal-line semimetals \cite{Nodal, PbTaSe2, ZrSiSNeupane, ZrSiSSchoop} and beyond \cite{Bradlyn, CoSi, RhSi, Co2MnGa}. Different from the point degeneracy at the topological band crossings between the bulk valence and conduction bands in other classes of TSM, the nodal-line semimetals feature band crossing that extends along a line or a loop \cite{Nodal}. Among the nodal-line semimetals, $\mathrm{ZrSiS}$ and similar $\mathrm{111}$ materials have been the most studied ones because of their ability to host multiple fermionic states including the nodal-line fermion and non-symmorphic fermion originating from the square net of Group-IV elements \cite{ZrSiSNeupane, ZrSiSSchoop, Topp, Hu, Takane, ZrSiX, Chen, Lou, ZrGeTe, Fu}. In addition to the topological fermions, these materials may also exhibit exotic properties including flat optical conductivity, unconventional magnetotransport, and unconventional mass enhancement \cite{Schilling, Ali, Lv, Singha, Kumar, Pezzini}.\\

Lanthanide ($Ln$)-based $Ln\mathrm{SbTe}$ systems belonging to the $\mathrm{ZrSiS}$-type family of materials have recently become attractive to researchers because of (a) Sb-square net, less studied compared to Si-square net materials, (b) possible magnetic ordering from the $Ln$ magnetic moments, and (c) possible correlation effects brought by the $4f$ electrons of $Ln$ elements. Despite these attractions, the studies on the electronic structure of these $Ln\mathrm{SbTe}$ systems have been limited. One of the earlier works on one of such systems - antiferromagnetic (AFM) $\mathrm{GdSbTe}$ with Neel temperature of about $\mathrm{12~K}$ \cite{GdSbTeSankar} - revealed a nodal-line state as well as an AFM Dirac state protected by the combination of broken time-reversal symmetry and rotoinversion symmetry \cite{GdSbTe}. $\mathrm{CeSbTe}$ has been shown to be a potential ground for several topological features \cite{CeSbTe1} and also a more symmetric non-symmorphic Dirac state due to higher spin-orbit coupling (SOC) than $\mathrm{ZrSiS}$ \cite{CeSbTe2}. While $\mathrm{HoSbTe}$ \cite{HoSbTe1, HoSbTe2} and $\mathrm{DySbTe}$ \cite{DySbTe} exhibit SOC gap along all high-symmetry directions leading to weak topological insulating state \cite{DySbTe, HoSbTe3}, $\mathrm{LaSbTe}$ features gapless nodal-line that persists even with SOC included \cite{LaSbTe}. Previous work on $\mathrm{SmSbTe}$ revealed multiple Dirac nodes that correspond to gapless nodal-line in this material system \cite{SmSbTe1}. Presence of Dirac nodal-lines in this system was also reported in another study, which also demonstrated the possibility of Kondo effects and electronic correlation enhancements \cite{SmSbTe2}. These results demonstrate that the electronic structure of $Ln\mathrm{SbTe}$ depends on the choice of $Ln$ element, however, a comprehensive understanding of how the electronic structure and topology in this family of materials evolve demands more electronic structure studies of $Ln\mathrm{SbTe}$ systems with different $Ln$ elements. Out of such systems, $\mathrm{NdSbTe}$ has been reported to feature metamagnetic transitions and potentially a coexisting Kondo localization \cite{NdSbTe1}, which has been reported in $\mathrm{CeSbTe}$ as well \cite{CeSbTe3, CeSbTe4, CeSbTe5}. Although transport and magnetic measurements have been performed in $\mathrm{NdSbTe}$ \cite{NdSbTe1, NdSbTe2}, the experimental measurement of the electronic structure is yet to be reported.\\

\begin{figure*} [ht]
\includegraphics[width=1\textwidth]{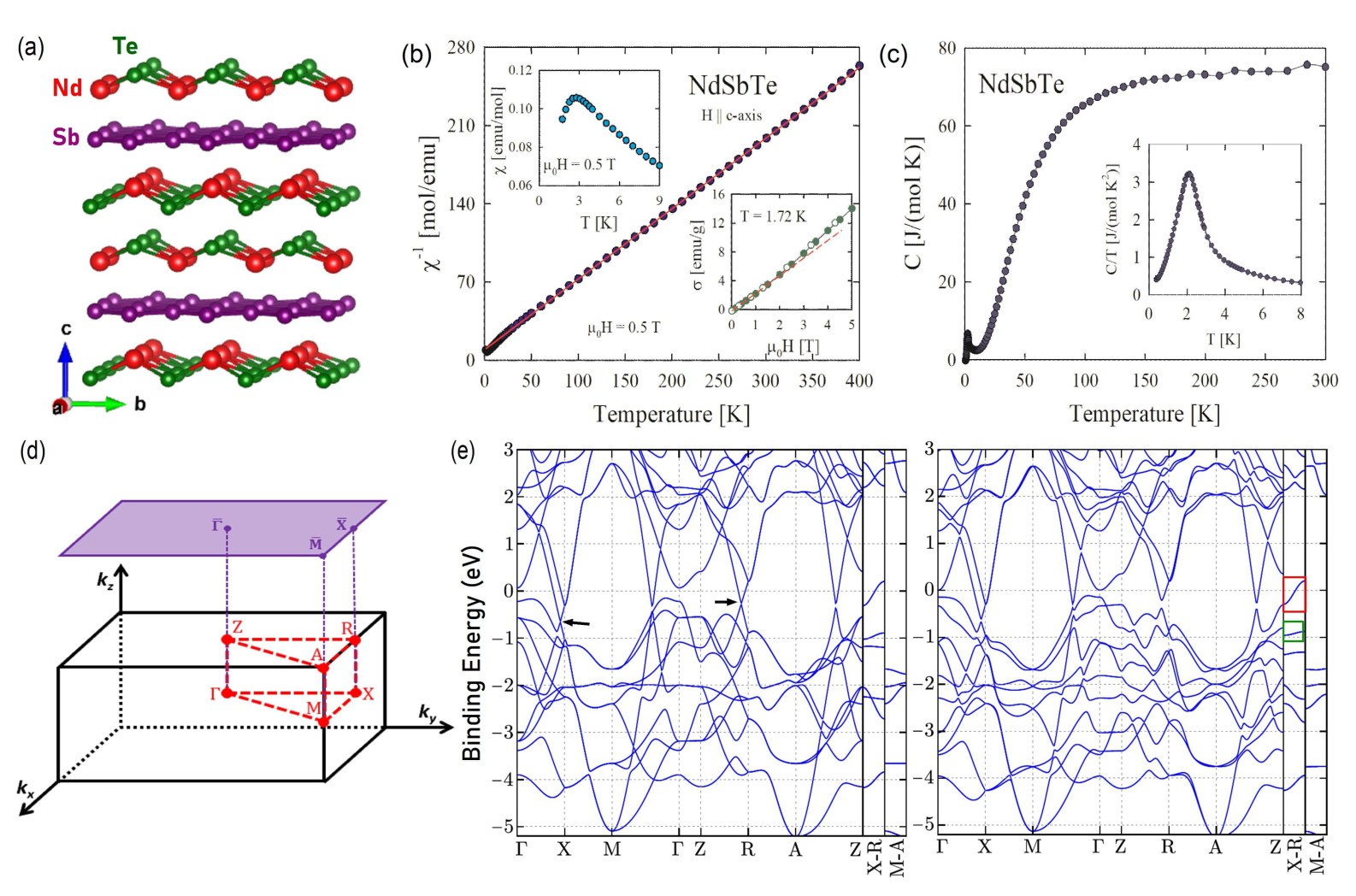}
\caption{Bulk properties of $\mathrm{NdSbTe}$. (a) Nonsymmorphic tetragonal crystal structure of $\mathrm{NdSbTe}$. (b) Temperature dependence of the inverse magnetic susceptibility measured in a magnetic field of $\mathrm{0.5~T}$ applied along the tetragonal axis. Solid red line represents the Curie-Weiss fit. Top left inset: low-temperature magnetic susceptibility data. Bottom right inset: magnetic field variation of the magnetization taken at $\mathrm{1.72~K}$ with increasing (solid symbols) and decreasing (open symbols) field strength. Dashed line emphasizes a linear behavior below the metamagnetic transition. (c) Temperature variation of the specific heat. Inset: low-temperature data plotted as the specific heat over temperature ratio versus temperature. (d) Bulk Brillouin zone and its projection onto $\mathrm{(001)}$ surface Brillouin zone. High symmetry points are marked. (e) Left: Calculated bulk bands along various high-symmetry directions without considering the spin-orbit coupling. Right: Calculated bulk bands with spin-orbit coupling taken into account. }
\label{F1}
\end{figure*}

In this paper, we report a detailed study of $\mathrm{NdSbTe}$ through high-resolution angle-resolved photoemission spectroscopy (ARPES) measurements complemented by first-principles calculations and thermodynamic characterizations. Thermodynamic measurements show that $\mathrm{NdSbTe}$ is AFM with a transition temperature of around $\mathrm{2~K}$. Our experimental data in the paramagnetic phase shows the presence of gapless nodal-lines along multiple high-symmetry directions in agreement with the predictions from the first-principles calculations. The bands associated with the nodal-line around the $\mathrm{\Gamma}$ point across the $\mathrm{\Gamma}-\mathrm{M}$ direction are observed to be sensitive to light polarization. Couple of nodal-lines are observed along the $\mathrm{X}-\mathrm{R}$ direction, with one in the vicinity of the Fermi level and the other residing around $\mathrm{800~meV}$ below it. Overall, this study reveals the electronic structure and topological states in $\mathrm{NdSbTe}$, for the first time, and presents a valuable platform towards having an understanding of the interplay of topology and $Ln$ elements in the  $Ln\mathrm{SbTe}$ family of quantum materials.

\begin{figure*} [ht]
\includegraphics[width=1\textwidth]{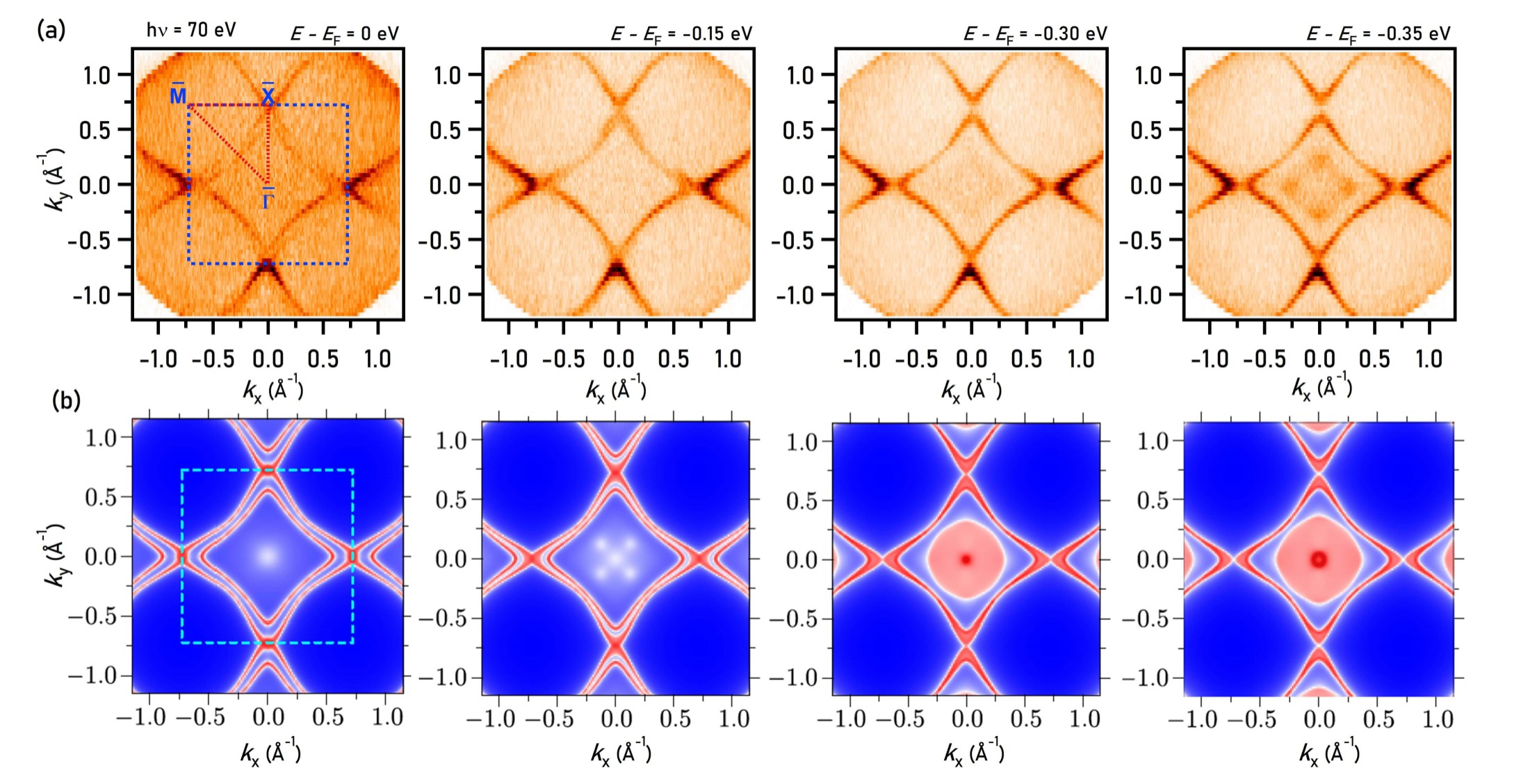}
\caption{Fermi surface and constant energy contours. (a) ARPES measured Fermi surface (first panel) and constant energy contours at various binding energies as noted on top of each plots. (b) Respective Fermi surface and energy contours obtained from first-principles calculations.}
\label{F2}
\end{figure*}

\begin{center} \textbf{II. METHODS} \end{center}
High-quality single crystals of $\mathrm{NdSbTe}$ were synthesized using chemical vapor transport technique and characterized using energy dispersive X-ray spectroscopy and single crystal X-ray diffraction. Thermodynamic (heat capacity, magnetic susceptibility, magnetization) measurements were performed using commercial equipment. ARPES experiments were carried out at the Stanford synchrotron radiation lightsource endstation 5-2  at a temperature of $18~ \mathrm{K}$. The energy resolution of the ARPES analyzer was set better than 20 meV.  Density-functional theory \cite{HK64,KS65} based first-principles calculations were carried out using the Vienna ab initio simulation package \cite{KresseHafner94, KresseFurthmuller96, KresseJoubert99} on the basis of projector augmented wave potential \cite{Blochl94}. For more details, see Section 1 in the supplemental material (SM) \cite{SM}.

\begin{center} \textbf{III. RESULTS} \end{center}
\noindent\textbf{A. Bulk properties of $\mathrm{NdSbTe}$}\\
The crystal structure of $\mathrm{NdSbTe}$ is presented in Fig.~\ref{F1}(a), where $\mathrm{Nd-Te}$ atomic layers sandwich the square nets formed by the $\mathrm{Sb}$ atoms. The structure is tetragonal in the nonsymmorphic space group number 129 ($P4/nmm$). Figure~\ref{F1}(b) depicts the temperature dependence of the inverse magnetic susceptibility of $\mathrm{NdSbTe}$ measured in a magnetic field applied along the crystallographic c-axis. Above $\mathrm{50~K}$, $\chi^{-1}(T)$ can be approximated by the Curie-Weiss (CW) law (note the solid straight line) with the effective magnetic moment $\mu\mathrm{_{eff} = 3.57(1)~\mu_B}$ and the paramagnetic Curie temperature $\theta\mathrm{_p = -17.4(2)~K}$. The value of $\mu\mathrm{_{eff}}$ is close to the theoretical Russell-Saunders prediction for trivalent $\mathrm{Nd}$ ion ($\mathrm{3.62~\mu_B}$). The negative sign of $\theta\mathrm{_p}$ signals AFM exchange interactions. Below $\mathrm{50~K}$, $\chi^{-1}(T)$ deviates from the CW behavior, likely due to the crystalline electric field (CEF) effect. As displayed in the top left inset of Fig.~\ref{F1}(b), $\chi(T)$ forms a maximum at $T\mathrm{_N = 2.7~K}$ that signals the long-range AFM ordering. Another indication of the AFM character of the magnetically ordered state is the behavior of the magnetization measured at $T\mathrm{ = 1.72~K}$ as a function of magnetic field. As can be inferred from the lower inset to Fig.~\ref{F1}(b), $\sigma(H)$ is initially linear and exhibits a metamagnetic-like transition near $\mathrm{2~T}$. In stronger fields, the magnetization shows a faint convex curvature and remains far from saturation. In a magnetic field of $\mathrm{5~T}$, the magnetization attains a value of $\mathrm{14.2(1)~emu/g}$ that corresponds to the magnetic moment of $\mathrm{1~\mu_B}$ that is much smaller than the theoretical prediction for $\mathrm{Nd^{3+}}$ ion ($\mathrm{3.27~\mu_B}$). This reduction results mostly from the CEF interaction and single-ion anisotropy. The $\sigma(H)$ variation exhibits no hysteresis effect, in concert with the AFM nature of the magnetic ground state of $\mathrm{NdSbTe}$. Figure~\ref{F1}(c) presents the temperature dependence of the specific heat of $\mathrm{NdSbTe}$. Close to room temperature, $C(T)$ attains a value of about $\mathrm{74~J/(mol~K)}$ that is nearly equal to the Dulong-Petit limit ($\mathrm{74.8~J/(mol~K)}$). With decreasing temperature, the specific heat decreases in an usual manner. The AFM phase transition manifests itself as a distinct mean-field-like anomaly in $C(T)$. As clearly seen in the inset to Fig.~\ref{F1}(c), the latter feature exhibits an extended tail above $T\mathrm{_N}$ that can be attributed to short-range exchange interactions. It is worth noting that overall the thermodynamic data collected for the single crystals of $\mathrm{NdSbTe}$ used in our ARPES study is very similar to the results obtained before by other research groups \cite{NdSbTe1, NdSbTe2}.\\

\begin{figure*} [ht!]
\includegraphics[width=1\textwidth]{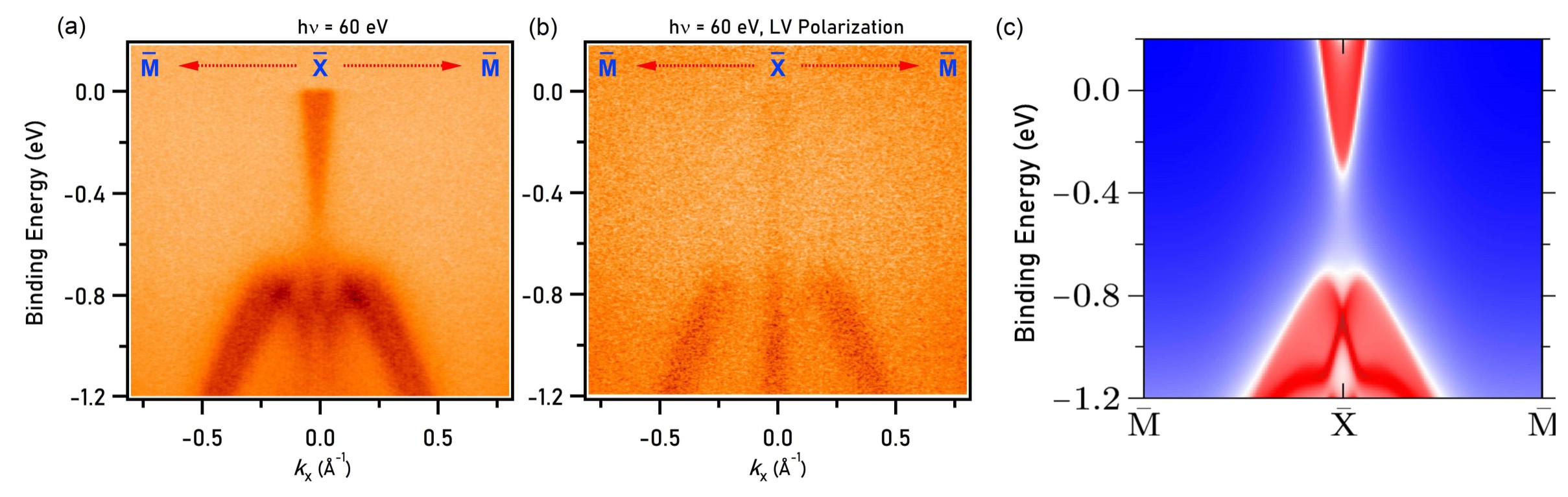}
\caption{ Surface state along $\mathrm{\overline{M}}-\mathrm{\overline{X}}-\mathrm{\overline{M}}$. (a)-(b) Dispersion maps along $\mathrm{\overline{M}}-\mathrm{\overline{X}}-\mathrm{\overline{M}}$ measured with linear horizontal and linear vertical polarized light with energy $\mathrm{60~eV}$, respectively. (c) Calculated surface spectrum along $\mathrm{\overline{M}}-\mathrm{\overline{X}}-\mathrm{\overline{M}}$. }
\label{F3}
\end{figure*}

The lattice parameters of $\mathrm{NdSbTe}$ were theoretically optimized, and the so-obtained values $a = b = \mathrm{4.371~\AA}$, and $c = \mathrm{9.457~\AA}$ are  close to those reported before ($a = b = \mathrm{4.312~\AA}$, and $c = \mathrm{9.371~\AA}$ \cite{Charvillat}). The Wyckoff positions occupied by $\mathrm{Nd}$, $\mathrm{Sb}$, and $\mathrm{Te}$ atoms are respectively: $2c$ ($1/4$,$1/4$,$0.2769$), $2a$ ($3/4$,$1/4$,$0$), and $2c$ ($1/4$,$1/4$,$0.6279$). In Fig. \ref{F1}(d), we present the three-dimensional bulk Brillouin zone and  its projection onto the two-dimensional (001) surface Brillouin zone. The high-symmetry points are marked on both the bulk and surface Brillouin zones. The left and the right panels in Fig.~\ref{F1}(e) show the calculated bulk bands along different high-symmetry directions without and with the consideration of SOC, respectively.  Dirac nodal-lines are realized along the $\mathrm{X}-\mathrm{R}$ and $\mathrm{M}-\mathrm{A}$ directions (due to $P4/nmm$ symmetry). Within $\mathrm{\pm1~eV}$ of the Fermi level, the band structure without SOC shows Dirac nodes at the X and R points that correspond to nodal-lines along the $\mathrm{X}-\mathrm{R}$ direction. Dirac nodes are also present along $\mathrm{\Gamma}-\mathrm{X}$ and $\mathrm{Z}-\mathrm{R}$ (shown by black arrows in Fig.~\ref{F1}(e)). However, when SOC is considered, these Dirac nodes along the $\mathrm{\Gamma}-\mathrm{X}$ and $\mathrm{Z}-\mathrm{R}$ directions gap out. These connect to the gapped nodal-line  along $\mathrm{\Gamma}-\mathrm{M}$ (or $\mathrm{Z}-\mathrm{A}$) forming the diamond-like structure (see SM Fig. S2 for the exact bulk gap in the diamond-like structure \cite{SM}). The Dirac nodes at the $\mathrm{X}$ and $\mathrm{R}$ points remain gapless leading to gapless nodal-lines along $\mathrm{X}-\mathrm{R}$, even with SOC considered (also see Fig. S2 and Section 2 in the SM \cite{SM}). Specifically, the nodal-line near the Fermi level (enclosed in red rectangle) is dispersive with Dirac node below the Fermi level at the $\mathrm{X}$ point, which moves upwards and above the Fermi level on reaching the $\mathrm{R}$ point. The other nodal line that lies slightly above $\mathrm{-1~eV}$ (enclosed in green rectangle) shows very little dispersion along the $\mathrm{X}-\mathrm{R}$ direction. \\

\begin{figure*} [ht!]
\includegraphics[width=1\textwidth]{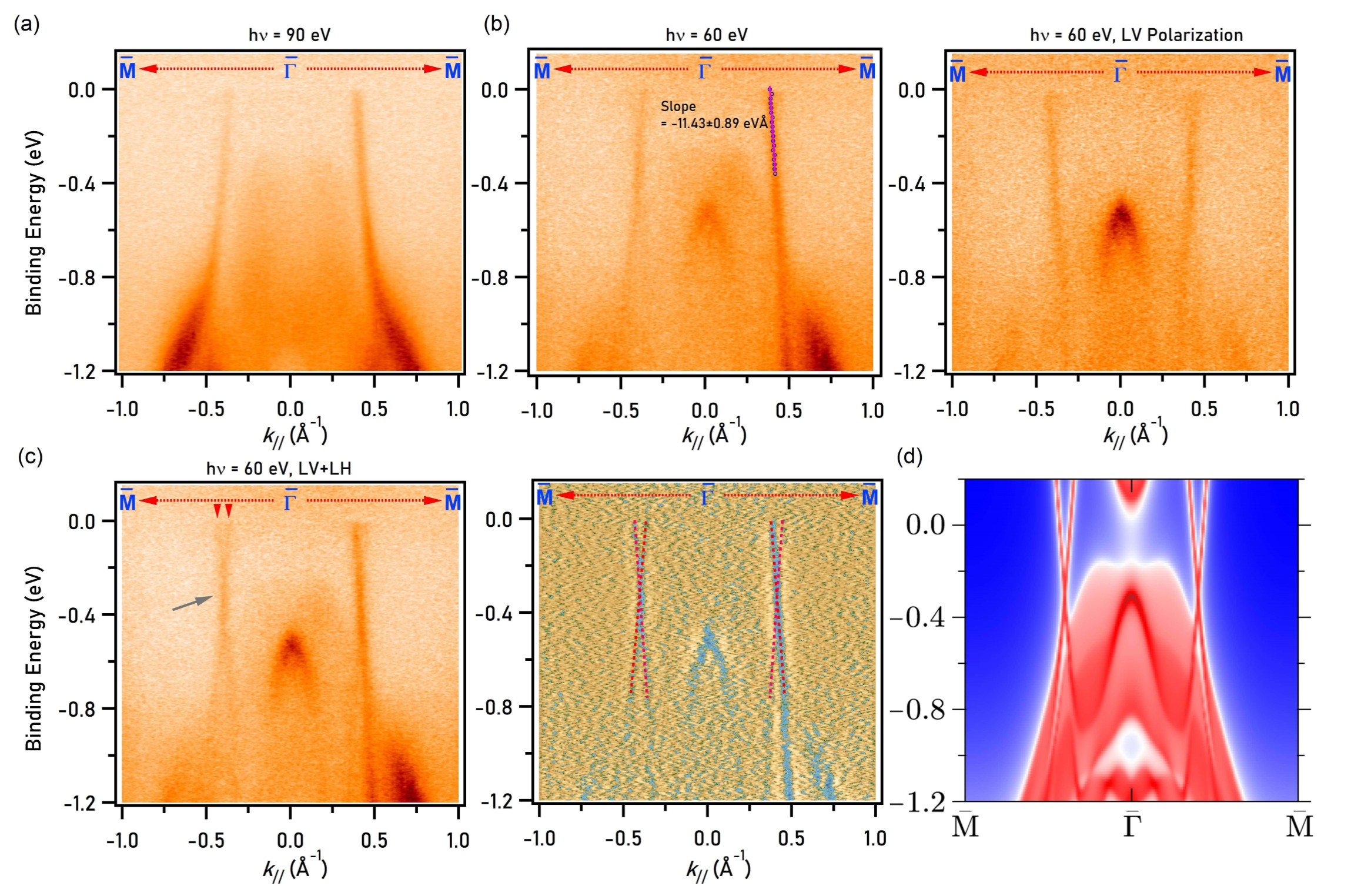}
\caption{Band dispersion along $\mathrm{\overline{M}}-\mathrm{\overline{\Gamma}}-\mathrm{\overline{M}}$. (a)-(b) Dispersion maps along the $\mathrm{\overline{M}}-\mathrm{\overline{\Gamma}}-\mathrm{\overline{M}}$  direction measured at photon energies of $\mathrm{90~eV}$ and $\mathrm{60~eV}$, respectively. Left panel in (b) shows the dispersion map measured with linear vertical polarization. (c) Dispersion map at $\mathrm{60~eV}$ photon energy with both linear horizontal and liner vertical data added. Right panel shows the second derivative plot of the dispersion map in the left panel. (d) Calculated surface band spectrum along $\mathrm{\overline{M}}-\mathrm{\overline{\Gamma}}-\mathrm{\overline{M}}$.}
\label{F4}
\end{figure*}

\begin{figure*} [ht!]
\includegraphics[width=1\textwidth]{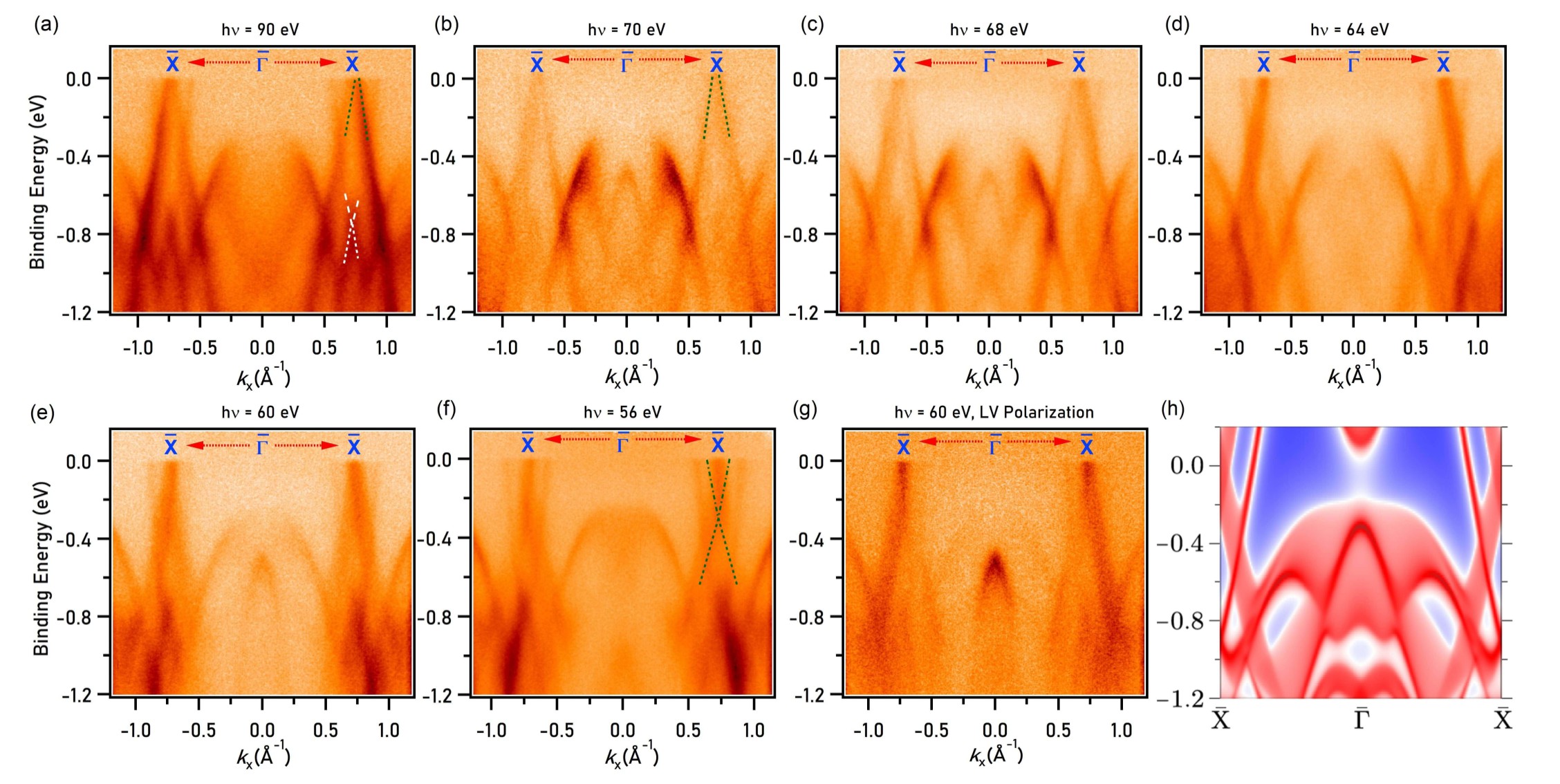}
\caption{Band dispersion along $\mathrm{\overline{X}}-\mathrm{\overline{\Gamma}}-\mathrm{\overline{X}}$. (a) Dispersion map along the $\mathrm{\overline{X}}-\mathrm{\overline{\Gamma}}-\mathrm{\overline{X}}$  direction measured with different photon energy of $\mathrm{90~eV}$. (b) Photon energy dependent dispersion maps along $\mathrm{\overline{X}}-\mathrm{\overline{\Gamma}}-\mathrm{\overline{X}}$ with the value of photon energy noted on top of each plot. (c)  $\mathrm{\overline{X}}-\mathrm{\overline{\Gamma}}-\mathrm{\overline{X}}$ band structure measured with $\mathrm{60~eV}$ linear vertical polarized photon source. (d) Calculated surface projected band structure along the $\mathrm{\overline{X}}-\mathrm{\overline{\Gamma}}-\mathrm{\overline{X}}$ direction.}
\label{F5}
\end{figure*}

\noindent\textbf{B. ARPES measured energy contours and surface state at the $\mathrm{\overline{\textbf{X}}}$ point.}\\
Figures~\ref{F2}-\ref{F5} present the results of the ARPES measurements on the $\mathrm{NdSbTe}$ single crystal. Note that the light source is linear horizontal (LH) polarized unless specified on top of the figure. The ARPES measured Fermi surface (FS) in Fig.~\ref{F2}(a) (first panel) is similar to the $\mathrm{ZrSiS}$-type materials with a diamond shaped pocket centered at the $\mathrm{\overline{\Gamma}}$ point. The diamond pocket has double sheets, out of which the inner sheet is strongly visible and the outer one has very faint intensity. The double sheet nature of the FS is well reproduced in the  calculated FS (Fig.~\ref{F2}(b), first panel). The sheet separation in the momentum space reduces with increasing binding energy and finally turns into a single sheet at higher values of binding energies as seen in the constant energy contours at  $\mathrm{300~meV}$ binding energy, where a new pocket emerges at the center of the Brillouin zone. Several bulk pockets appear at higher binding energies, which can be clearly visualized from the constant energy contour plots at binding energies of $\mathrm{350~meV}$, $\mathrm{500~meV}$, and $\mathrm{700~meV}$ (see section 2 in the SM \cite{SM}). The calculated energy contours well reproduce the experimental ones. The exception of the intense circular pocket at higher binding energy can be explained from the polarization dependence, which can be seen from the polarization dependent maps presented in section 3 of the SM  \cite{SM}. \\

In order to understand the underlying electronic structure of $\mathrm{NdSbTe}$ at low temperatures, we analyze the dispersion maps along various high-symmetry directions in Figs.~\ref{F3}-\ref{F5}. We begin with the dispersion maps along the $\mathrm{\overline{M}}-\mathrm{\overline{X}}-\mathrm{\overline{M}}$, which is presented in Fig.~\ref{F3}. A gapped Dirac-like state is observed at the $\mathrm{\overline{X}}$ point. The dispersion of this state does not seem to depend on the choice of photon energy (see section 5 in the SM \cite{SM} for more photon energy measurements), which implies that this state is surface originated. The surface state co-exists with bulk states (see slab calculation in the SM section 4 \cite{SM}). As seen in Fig.~\ref{F3}(a), the  surface state is strong in measurement with LH polarized photon source and strongly suppressed in measurement with linear vertical (LV) polarized light source (Fig.~\ref{F3}(b)). The calculated surface spectrum is presented in Fig.~\ref{F3}(c), which fairly reproduces the experimental observation of the gapped surface state at the $\mathrm{\overline{X}}$ point. \\

\noindent\textbf{C. Observation of gapless nodal-line states}\\
Next, we move on to the experimental electronic band dispersion along the  $\mathrm{\overline{M}}-\mathrm{\overline{\Gamma}}-\mathrm{\overline{M}}$ direction. Figure~\ref{F4}(a) represents a dispersion map measured with photon energy of $\mathrm{90~eV}$ along this direction. A single band can be seen crossing the Fermi level. Similar band dispersion is observed in measurement carried out with $\mathrm{60~eV}$ photon energy (Fig.~\ref{F4}(b), left panel). This band has almost linear dispersion up to $\mathrm{\sim0.8~eV}$ below the Fermi level and is slightly converging towards the $\mathrm{\overline{\Gamma}}$ point. The Fermi velocity is very large ($\mathrm{11.43\pm0.89~eV\AA}$ as obtained from the slope of the linear fit of peaks in the momentum distribution curves within $\mathrm{-360~meV}$ binding energy, see Fig.~\ref{F4}(b) left panel) compared to the $\mathrm{\overline{\Gamma}}-\mathrm{\overline{M}}$ bands in $\mathrm{ZrSiS}$ \cite{ZrSiSvf}. The observation of a double sheet FS means that there should be two bands crossing the Fermi level along this direction. Strong suppression of outer sheet in the FS map and the observation of single band extending upto the Fermi level along $\mathrm{\overline{M}}-\mathrm{\overline{\Gamma}}-\mathrm{\overline{M}}$ in the measurements with LH polarization pointed to the possibility of polarization dependence of the outer band. We, therefore, performed measurement with LV polarized light (Fig.~\ref{F4}(b), right panel), where we again observed a single band crossing the Fermi level. This band is also almost linear above $\mathrm{~-500~meV}$. Interestingly, unlike the band observed in LH polarization, this band seems to slightly diverge away from the $\mathrm{\overline{\Gamma}}$ point. In Fig.~\ref{F4}(c), we present the dispersion map along $\mathrm{\overline{M}}-\mathrm{\overline{\Gamma}}-\mathrm{\overline{M}}$ with the LH and LV polarization matrices added. We can clearly observe two bands crossing the Fermi level, out of which the inner one is sensitive to LH polarization and the outer one is sensitive to LV polarization (also see section 3 in the SM \cite{SM}). This  explains the double sheet nature of the diamond pocket in the FS. These two bands cross each other at around $\mathrm{300~meV}$ below the Fermi level and disperse almost linearly across this crossing point giving an impression of a gapless nodal-line along $\mathrm{\overline{\Gamma}}-\mathrm{\overline{M}}$ direction. This can be clearly visualized in the second derivative plot presented in Fig.~\ref{F4}(c) (right panel), where the crossing between the bands is shown by two red-colored dashed lines. The calculated band structure along this direction is presented in Fig.~\ref{F4}(d), which clearly shows this crossing feature. Despite the experimental observation of the gapless nodal-line in experiments, one important thing to note is that the bulk calculation shows gapped nature along this direction (Fig.~\ref{F1}(d)) and the calculation in Fig.~\ref{F4}(d) is integrated for all values of $k_z$. The $k_z$ plane dependent bulk bands presented in section 6 of the SM \cite{SM} also shows a gapped nature for any particular value of $k_z$, which is undetectable in ARPES data.  {Such a nodal-line with undetectable gap within the experimental resolution was reported in a similar $Ln\mathrm{SbTe}$-type material \cite{SmSbTe2}.} The ARPES measurement in the vacuum ultraviolet region has poor $k_z$ resolution and hence the data observed in the experiments probably covers a certain range of $k_z$ value rather than a certain $k_z$ plane. The other possibility could be because of the discrepancy between experimental data and theoretical calculations that is usual for metallic/semimetallic systems. \\

In Fig.~\ref{F5}, we present the dispersion maps along the $\mathrm{\overline{X}}-\mathrm{\overline{\Gamma}}-\mathrm{\overline{X}}$ direction. The measured dispersion map along this direction with a photon energy of $\mathrm{90~eV}$ (Fig.~\ref{F5}(a)) shows a linearly dispersing  band (green dashed line) crossing the Fermi level around the $\mathrm{\overline{X}}$ point. In addition, another linear band touching is present at the $\mathrm{\overline{X}}$ point about $\mathrm{760~meV}$ below the Fermi level. In Figs.~\ref{F5}(b-f), $\mathrm{\overline{X}}-\mathrm{\overline{\Gamma}}-\mathrm{\overline{X}}$ band dispersion measured with different photon energies are presented. The linear dispersion in the vicinity of the Fermi level seems to be strongly photon energy dependent. This photon energy dependence is indicative of the bulk origination of the bands associated with the linear dispersion. The crossing point of the linear dispersion lies well below the Fermi level at a photon energy of $\mathrm{56~eV}$ ($k_z \sim \mathrm{0}$, Fig.~\ref{F5}(f)) and it shifts upwards and above the Fermi level on increasing the photon energy up to $\mathrm{70~eV}$($k_z \sim \mathrm{\pi}$, Fig. \ref{F5}(b)). This agrees well with the theoretical prediction of the dispersive and gapless nodal-line along the $\mathrm{X}-\mathrm{R}$ direction. {The presence of gapless nodal-line in the vicinity of the Fermi level has been recently reported in similar compounds $\mathrm{SmSbTe}$  \cite{SmSbTe1} and $\mathrm{LaSbTe}$ \cite{LaSbTe}}. The band-touching feature around $\mathrm{760~meV}$, on the other hand, seems to have very little change with photon energy variation, in agreement with the almost dispersionless and gapless nodal-line along $\mathrm{X}-\mathrm{R}$ direction as seen in the bulk-band calculations. This feature is also observed to be sensitive to LH polarization and is strongly suppressed when the incident photon is LV polarized (see Fig.~\ref{F5}(g)). The observation of the Dirac nodes along the $\mathrm{\overline{X}}-\mathrm{\overline{\Gamma}}-\mathrm{\overline{X}}$ direction is well reproduced in the calculated surface electronic structure shown in Fig.~\ref{F5}(h) (also see section 6 in the SM \cite{SM}).

\begin{center} \textbf{III. CONCLUSION} \end{center}
In summary, through high-resolution ARPES measurements supported by theoretical calculations, we studied the detailed low-temperature electronic structure in $Ln$ based $\mathrm{ZrSiS}$-type material $\mathrm{NdSbTe}$. The thermodynamic measurements indicated an AFM transition in this material with Neel temperature, $T_N\mathrm{\sim2~K}$. Our ARPES results in the paramagnetic phase detect three gapless nodal lines in this material system - two along bulk $\mathrm{X}-\mathrm{R}$ direction and one around the $\mathrm{\Gamma}$ point formed by steep bands with high Fermi velocity that constitute the diamond shape. Although theory predicts that the nodal-line around the $\mathrm{\Gamma}$ point is gapped, experimental results show gapless feature. The theoretical calculations well reproduce the experimental results. This work presents $\mathrm{NdSbTe}$ as a multiple nodal-line fermion system and provides a new platform to understand the evolution of topology and electronic structure among the $Ln\mathrm{SbTe}$ family of materials.

\begin{center} \textbf{ACKNOWLEDGMENTS} \end{center}
M.N. acknowledges the support from the National Science Foundation under CAREER award DMR-1847962 and the Air Force Office of Scientific Research MURI Grant No. FA9550-20-1-0322. D.K. was supported by the National Science Centre (Poland) under research grant 2021/41/B/ST3/01141. A.P. acknowledges the support by National Science Centre (NCN, Poland) under Projects No. 2021/43/B/ST3/02166 and also appreciates the funding in the frame of scholarships of the Minister of Science and Higher Education (Poland) for outstanding young scientists (2019 edition, No. 818/STYP/14/2019).  The use of Stanford Synchrotron Radiation Lightsource (SSRL) in SLAC National Accelerator Laboratory is supported by the U.S. Department of Energy, Office of Science, Office of Basic Energy Sciences under Contract No. DE-AC02-76SF00515. We thank Makoto Hashimoto and Donghui Lu for the beamline assistance at SSRL endstation 5-2.

 \clearpage

\setcounter{equation}{0}
\renewcommand{\theequation}{S\arabic{equation}}
\setcounter{figure}{0}
\renewcommand{\thefigure}{S\arabic{figure}}
\setcounter{section}{0}
\renewcommand{\thesection}{S\Roman{section}}
\setcounter{table}{0}
\renewcommand{\thetable}{S\arabic{table}}
\begin{center}
 \textbf{\Large \underline{Supplemental Material}}\\[0.5cm]
\end{center}

\begin{figure*} [h!]
\includegraphics[width=1.0\textwidth]{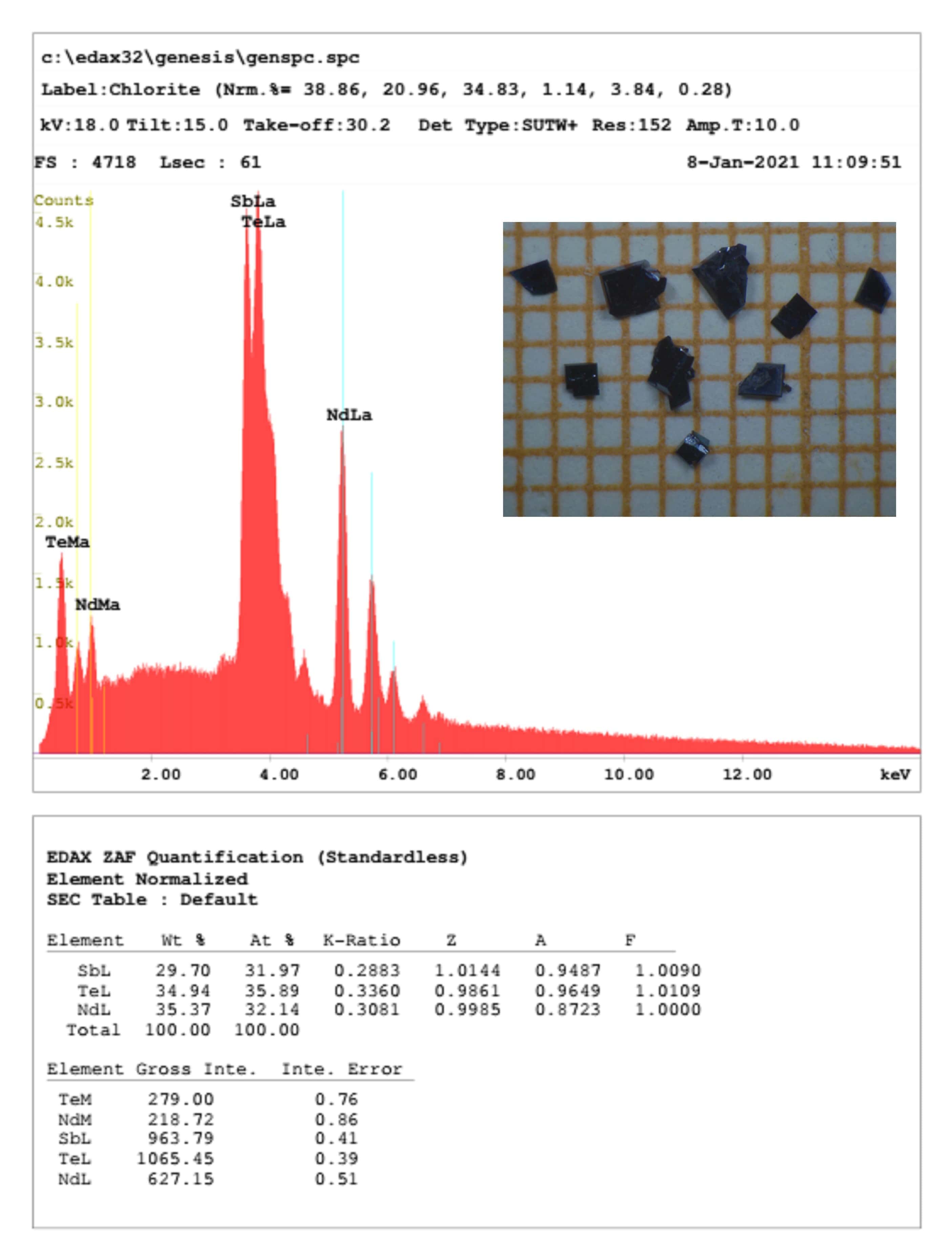}
\caption{Exemplary result of the microprobe analysis of the chemical composition of one of the $\mathrm{NdSbTe}$ crystals used in the physical properties measurements. Inset shows the single crystals of $\mathrm{NdSbTe}$ grown by chemical vapor transport method.}
\label{SF1}
\end{figure*}

 \begin{center}
\textbf{1. METHODS}\\
\textbf{1.1 Experimental techniques}\\
\textit{Crystal Structure and Sample Characterization} \end{center}
Synthesis of single-crystalline NdSbTe was carried out in two steps. First, a stoichiometric mixture of the elemental constituents (purity: $\mathrm{Nd ~–~ 99.9~ wt.\%}$, $\mathrm{Sb ~–~ 99.999~ wt.\%}$, $\mathrm{Te ~–~ 99.999 ~wt.\%}$) was placed in an evacuated quartz ampoule and slowly heated up to $\mathrm{1000~\degree C}$ in a resistance furnace, dwelled at this temperature for 7 days, and then cooled quickly to room temperature. The so-obtained polycrystalline material was utilized as a charge in single crystals growth performed by chemical vapor transport method with iodine as a transport agent. The evacuated quartz tube was placed in a two-zone furnace with the precursor hold at $\mathrm{950~\degree C}$ and the crystal growth taking place in a colder end of the tube maintained at $\mathrm{850~\degree C}$. After 20 days, black shiny crystals with platelet shape and dimensions up to $\mathrm{1 \times 1 \times 0.2~ mm^3}$ were obtained (see Fig. \ref{SF1} inset). They were found stable against air and moisture for few days only. 
    
Chemical composition and phase homogeneity of the prepared crystals were determined by energy-dispersive X-ray (EDX) analysis performed using a FEI scanning electron microscope equipped with an EDAX Genesis XM4 spectrometer. The EDX results proved single-phase character of the specimens and indicated the stoichiometry close to $\mathrm{1:1:1}$ (see Fig. \ref{SF1}).

Crystal symmetry of the obtained single crystals was examined on an Oxford Diffraction X'calibur four-circle single-crystal X-ray diffractometer equipped with a CCD Atlas detector. The experiment confirmed a tetragonal unit cell of the $\mathrm{ZrSiS}$-type with the lattice parameters close to those reported in the literature \cite{NdSbTe1S, NdSbTe2S}. Crystallinity and crystallographic orientation of the crystals selected for physical measurements were determined by means of Laue X-ray backscattering technique implemented in a LAUE-COS (Proto) system.\\

\begin{center} 
\textit{Thermodynamic properties Measurements}
\end{center}
Magnetic properties were investigated in the temperature range $\mathrm{1.72-400 ~K}$ and in magnetic fields up to $\mathrm{5~ T}$ using a Quantum Design MPMS-5 superconducting quantum interference device (SQUID) magnetometer. Heat capacity was measured in the temperature interval $\mathrm{0.37-300 ~K}$ employing the relaxation technique and $\mathrm{two-\tau}$ model implemented in a Quantum Design PPMS-9 platform equipped with a $\mathrm{^3He}$ refrigerator.\\

\begin{center}
\textit{Spectroscopic Characterization}
\end{center}

Spectroscopic characterization of the electronic band structure using ARPES was performed at the Stanford Synchrotron Radiation Lightsource endstation 5-2 at a temperature of $\mathrm{18~K}$. The energy resolution of the experimental set up was set better than $\mathrm{0.02~eV}$. The samples mounted on copper posts were loaded and then cleaved inside the ARPES chamber maintained at ultra-high vacuum better than $\mathrm{10^{-10}~Torr}$ before carrying out the measurements. \\

\begin{center}
\textbf{1.2 Computational details}
\end{center}
The density-functional theory (DFT) \cite{HK64S, KS65S} based first-principles calculations were performed using projector augmented-wave (PAW) potentials \cite{Blochl94S} implemented in the Vienna Ab initio simulation package {\sc VASP} \cite{KresseHafner94S, KresseFurthmuller96S, KresseJoubert99S}. Calculations were made within the generalized gradient approximation (GGA) in the Perdew, Burke, and Ernzerhof (PBE) parameterization \cite{PerdewBurke96S}. The energy cutoff for the plane-wave expansion was set to $\mathrm{350~eV}$. The structural parameters (lattice constants and atomic positions) were theoretically optimized because of the lack of experimental data about $z_\mathrm{Nd}$ and $z_\mathrm{Te}$ values. The optimization was performed in the conventional unit cell using the $15 \times 15 \times 7$ $\textbf{k}$-point grid in the Monkhorst--Pack scheme~\cite{MonkhorstPack76S}. The convergence was checked for these values. The excellent agreement of the lattice constants suggests a realistic value of the atom position. The atomic position in other compounds with similar structure e.g., $\mathrm{Sm}\mathrm{Sb}_{0.93}\mathrm{Te}_{1.07}$, $z_\mathrm{Sm} = 0.7244$ (or $0.2756$) and $z_\mathrm{Te} = 0.3734$ (or $0.6266$) \cite{Pandey2022S}, for $\mathrm{HoSbTe}$, $z_\mathrm{Ho} = 0.2762$ and $z_\mathrm{Te} = 0.3761$ (or $0.6239$) \cite{Igor2022S}, for $\mathrm{TbSbTe}$ $z_\mathrm{Tb} = 0.2758$ and $z_\mathrm{Te} = 0.3756$ (or $0.6244$) \cite{Igor2022S}, is close to our theoretical result, i.e., $z_\mathrm{Nd} = 0.2769$ and $z_\mathrm{Te} = 0.6279$.  In the optimization procedure, we assumed the AFM order, while the $4f$ states of Nd were treated within the DFT+$U$ scheme introduced by Liechtenstein  {\it et al.} \cite{LiechtensteinAnisimov95S} (with $\mathrm{U = 4.0~eV}$ and $\mathrm{J = 0.5~eV)}$. As a break of the optimization loop, we took the condition with an energy difference of $\mathrm{10^{-6}~eV}$ and $\mathrm{10^{-8}~eV}$ for ionic and electronic degrees of freedom. \\

The electronic band structure was also evaluated within {\sc Quantum ESPRESSO}  \cite{GiannozziBaroni09S,GiannozziAndreussi17S, GiannozziBaseggio20S} with PAW GGA PBE pseudopotential included in {\sc PsLibrary} \cite{Corso14S}. Exact results of the DFT calculations (with $\mathrm{Nd}$ $4f$ states as core states) were used to find the tight binding model in the basis of the maximally localized Wannier orbitals \cite{MarzhariMostofi12S, MarzhariVanderbilt97S, SouzaMarzhari01S}. It was performed using the {\sc Wannier90} software \cite{MostofiYates08S, MostofiYates14S, PizziVitale20S}. In our calculations, we used the $8 \times 8 \times 6$ full ${\bm k}$-point DFT calculation, starting from the $d$ orbitals of $\mathrm{Nd}$, and $p$ orbitals for $\mathrm{Sb}$ and $\mathrm{Te}$. The wannierization procedure (i.e., procedure of the tight binding model finding) is based on the exact DFT band structure. Within this procedure, storage of each wavefunction for all ${\bm k}$-points are necessary. Next, the Bloch wavefunction from DFT are used in the construction of the “rotation” and overlap matrices. More technical details can be found in the documentation of {\sc wannier90} or Ref. \cite{MarzhariMostofi12S}. Finally, the found $22$-orbital tight binding model of the $\mathrm{NdSbTe}$ was used for the investigation of the surface Green's function for semi-infinite system \cite{SanchoSancho85} using {\sc WannierTools} \cite{WuZhang18S} software. Within this calculation, the tight binding models of the bulk system is used to perform the properties of the system with “edge” (e.g. surface). More detailed theoretical description can be found in Ref. \cite{SanchoSancho85S}.  In order to match the experimental results, the Fermi level was shifted upwards by $\mathrm{200~meV}$ in the calculations.\\

\begin{figure*} [h!]
\includegraphics[width=1\textwidth]{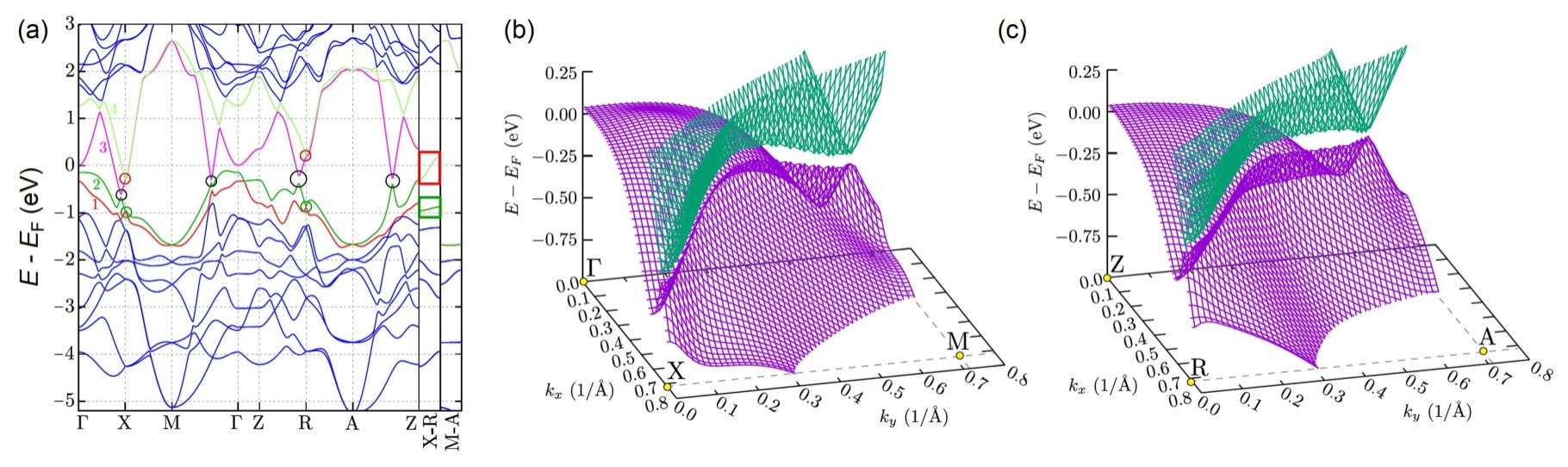}
\caption{Calculated bulk bands. (a) Bulk-band structure in the presence of spin-orbit coupling. (b-c) Gapped state formed by the bands 2 and 3 in (a) in the diamond-like structure for $k_z = 0$ (b) and $k_z = \pi$ (c).}
\label{SF2}
\end{figure*}

\begin{center}
\textbf{2. CALCULATED BULK BANDS INCLUDING SPIN-ORBIT COUPLING}
\end{center}
In \ref{SF2}(a), we reproduce the bulk band calculation including  spin-orbit coupling presented in main text Fig. 1(e) with four bands labeled 1 -- 4 in different colors. The bands 3 and 4 produce crossings at the $\mathrm{X}$ and $\mathrm{R}$ points (enclosed in red circles). These crossings produce a nodal-line along  $\mathrm{X}-\mathrm{R}$ in the vicinity of the Fermi level (enclosed in red colored rectangle). The crossing starts well below the Fermi level at the X point ($k_z = 0$) and shifts upward and  above the Fermi level on reaching the R point ($k_z = \pi$) giving rise to a nodal-line that is dispersive in energy axis. The bands 1 and 2 also produce crossings at the $\mathrm{X}$ and $\mathrm{R}$ points (enclosed in green circles), forming another nodal-line $\mathrm{X}-\mathrm{R}$ (enclosed in green rectangle). The bands 2 and 3 are responsible for the diamond shaped pockets seen in the FS and energy contours. These bands form a gapped state (enclosed in black circles) along $\mathrm{\Gamma}-\mathrm{X}$ and $\mathrm{\Gamma}-\mathrm{M}$ ($k_z = 0$) as well as $\mathrm{Z}-\mathrm{R}$ and $\mathrm{Z}-\mathrm{A}$ ($k_z = \pi$). The exact bulk gap in the diamond-like structure formed by bands 2 (upper) and 3(lower) can be seen in Figs. \ref{SF2}(b) and \ref{SF2}(c), respectively. \\

\begin{center}
\textbf{3. EXPERIMENTAL AND THEORETICAL ENERGY CONTOURS}
\end{center}
\begin{figure*} [h!]
\includegraphics[width=0.95\textwidth]{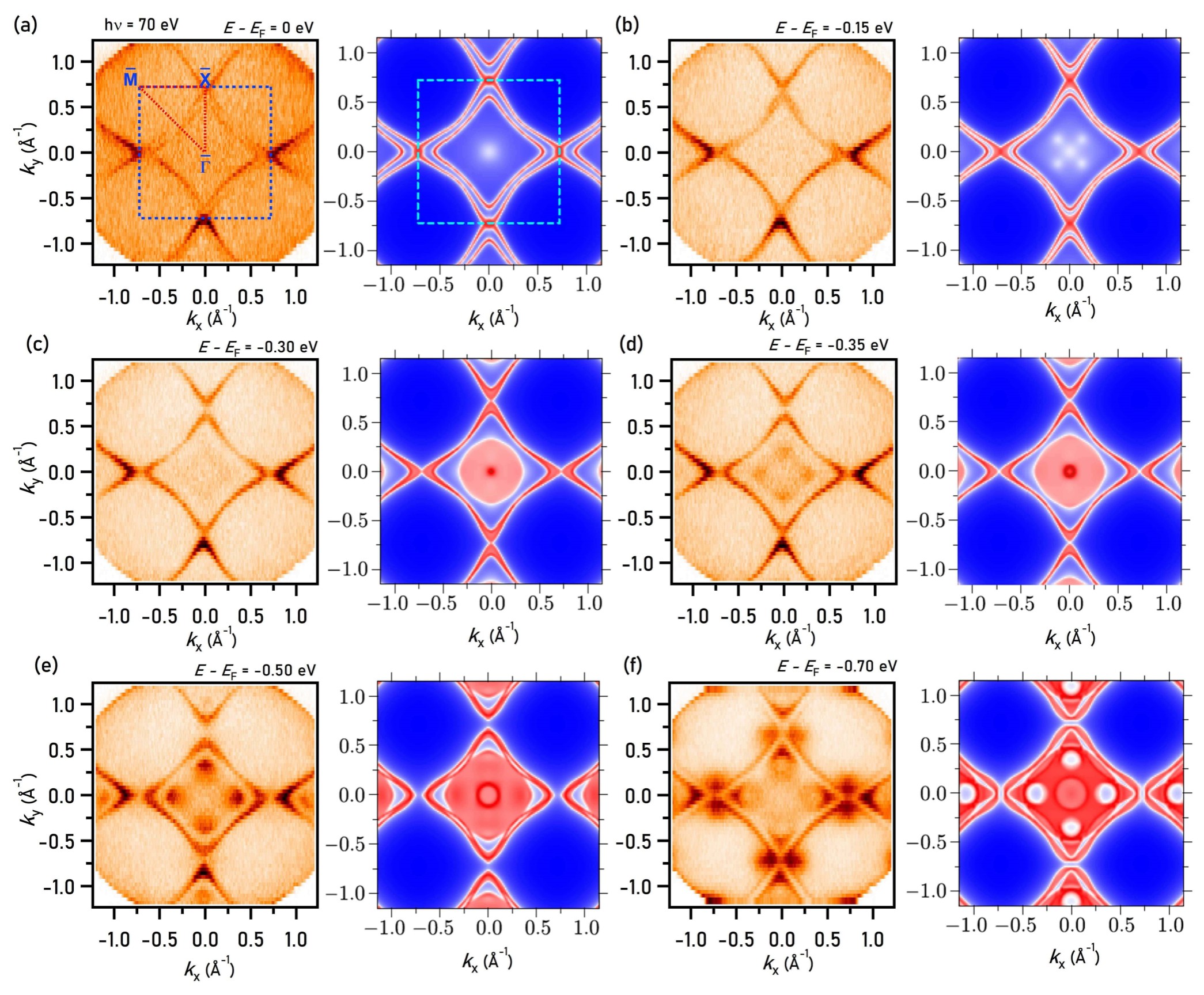}
\caption{FS and energy contours measured using a photon source of $\mathrm{90~eV}$. (a) FS and (b)-(f) energy contours at binding energy noted on top. The left panel in each plot represents the ARPES measured energy counter and the right panel represents the calculated energy contour.}
\label{SF3}
\end{figure*}
Figure \ref{SF3} presents the ARPES measured energy contours (left panels) and corresponding calculated energy contours (right panels) at the Fermi level and various binding energies as noted on top of each experimental plot. The measurements are done using $\mathrm{70~eV}$ photon energy. A typical $\mathrm{ZrSiS}$-type FS containing a $\mathrm{\Gamma}$ point centered diamond pocket is observed. The diamond pocket has double sheets (the inner one is intense and the outer one is faint because of polarization, see section 3 below), which is well reproduced in first-principles calculation. Double sheets merge towards each other and around $\mathrm{-300~meV}$  constant energy contour evolving into a single diamond pocket, where an extra inner pocket emerges. Several bulk pockets appear at higher binding energies, which can be clearly visualized from the constant energy contour plots at binding energies of $\mathrm{350~meV}$, $\mathrm{500~meV}$, and $\mathrm{700~meV}$. The experimental energy contours show fine agreement with the calculated ones. The exception of the intense circular pocket at higher binding energy can be explained from the polarization dependence (see section 4 below).\\

\begin{figure*} [h!]
\includegraphics[width=1.0\textwidth]{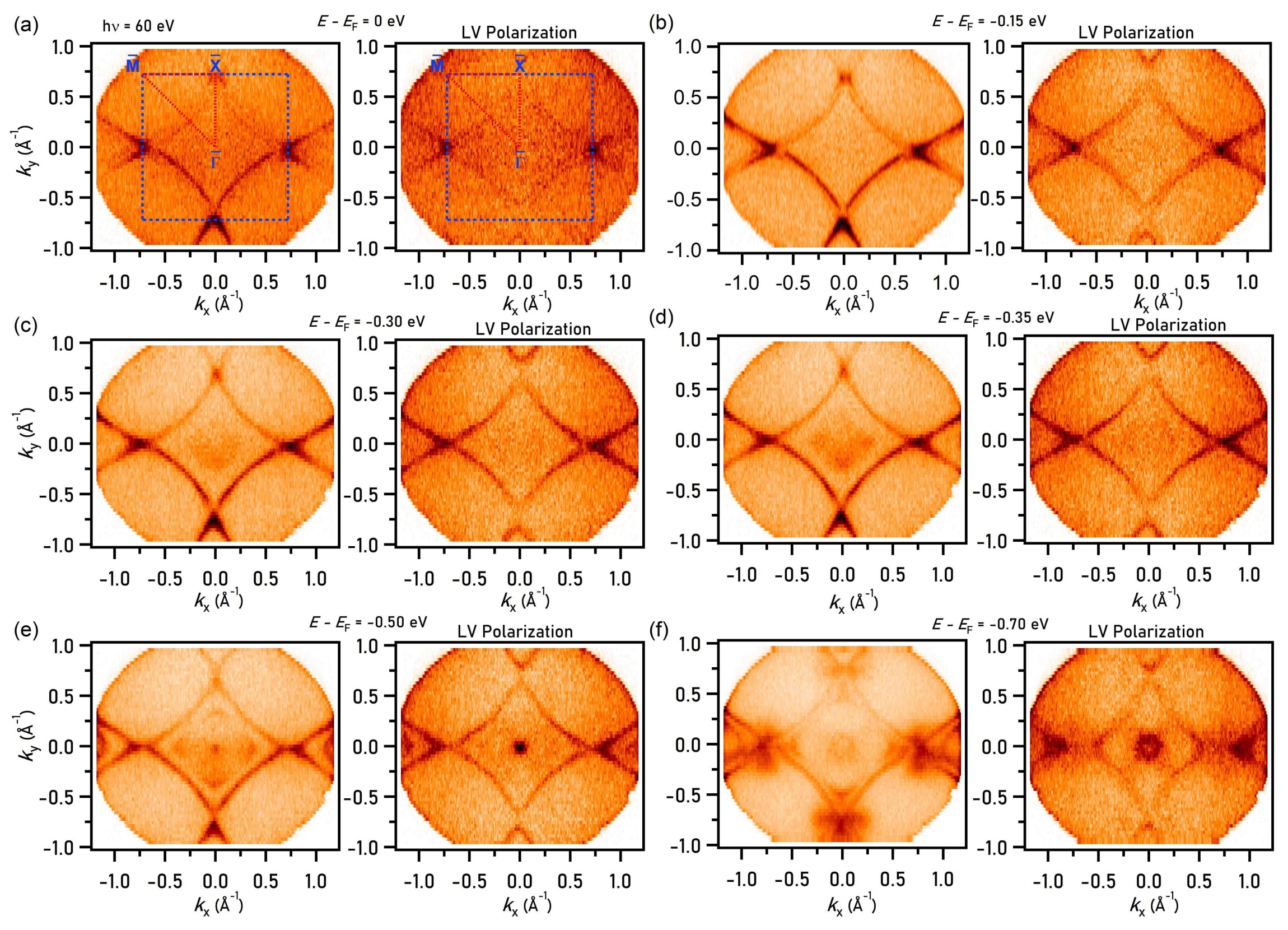}
\caption{FS and energy contours measured using a photon source of $\mathrm{60~eV}$. The left panels are measured using LH polarization and the right panels are measured using LV polarization. Corresponding binding energies are noted on top for each energy contours.}
\label{SF4}
\end{figure*}

\begin{center}
\textbf{4. FS AND ENERGY CONTOURS MEASURED WITH $\mathrm{60~eV}$  PHOTON ENERGY}
\end{center}
Figure \ref{SF4} shows the ARPES measured energy contours at the FS and various binding energies by using LH (left panels) and LV (right panels) polarized photon source of energy $\mathrm{60~eV}$. The FS contains a diamond-shaped pocket centered at the $\mathrm{\overline{\Gamma}}$ point. This pocket has a double sheet nature, where the sheets are sensitive to polarization of the incident photon. While the inner sheet is strong in LH polarized light, the outer sheet becomes prominent in LV polarized light. The sheets merge into a single-sheet on going down to higher binding energies creating a Dirac-like crossing obtained along the $\mathrm{\overline{M}}-\mathrm{\overline{\Gamma}}-\mathrm{\overline{M}}$ direction. Several bulk pockets emerge at higher binding energies, some of which show strong polarization dependence. For example, the hole pocket that appers at the center of surface Brillouin zone at a binding energy of $\mathrm{-500~meV}$ is more sensitive to LV polarization and the pockets emerging at the corner of the surface Brillouin zone at a binding energy of $\mathrm{-700meV}$ shows strong sensitivity to LH polarization.\\

\begin{figure*} [h]
\includegraphics[width=1\textwidth]{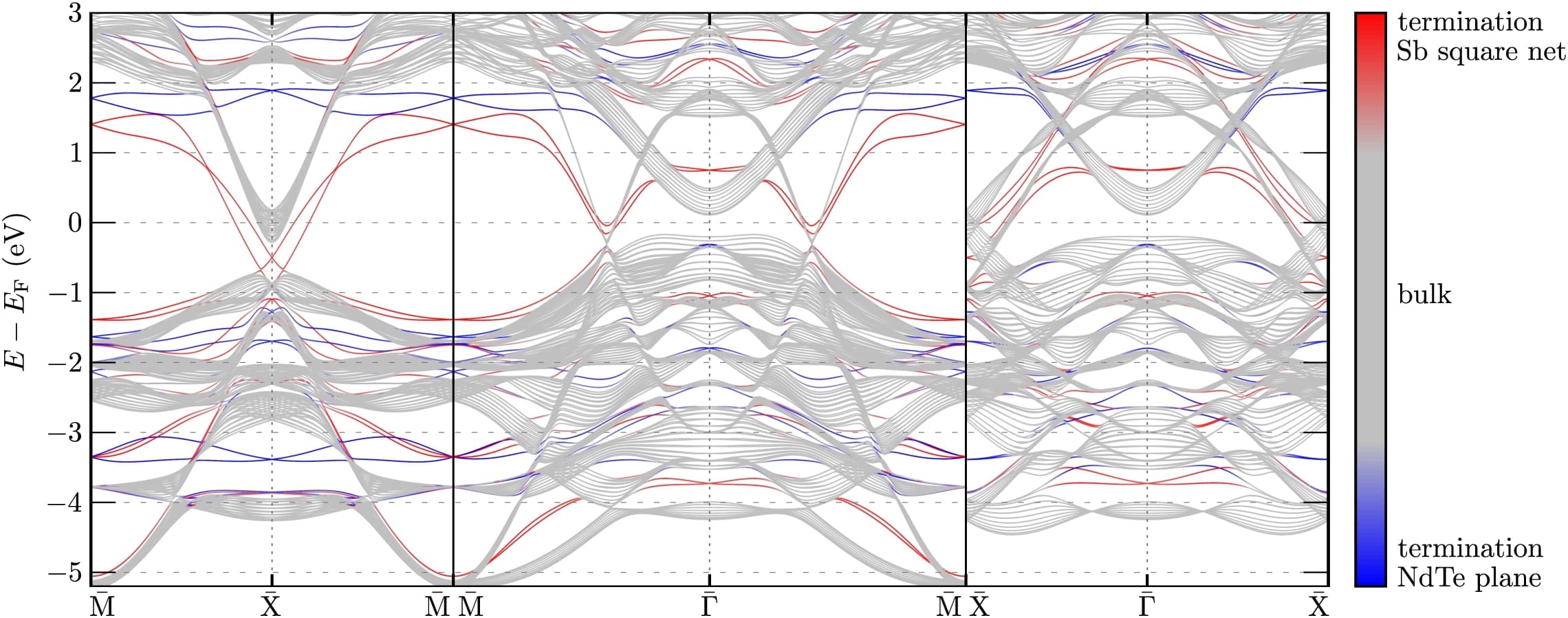}
\caption{Slab calculations along the $\mathrm{\overline{M}} - \mathrm{\overline{X}} - \mathrm{\overline{M}}$, $\mathrm{\overline{M}} - \mathrm{\overline{\Gamma}} - \mathrm{\overline{M}}$, and $\mathrm{\overline{X}} - \mathrm{\overline{\Gamma}} - \mathrm{\overline{X}}$ directions. Color of the line correspond to the surface termination as labeled. }
\label{SF5}
\end{figure*}

\begin{center}
\textbf{5. SLAB CALCULATIONS}
\end{center}
In Fig. \ref{SF5}, we present the slab calculations along the $\mathrm{\overline{M}} - \mathrm{\overline{X}} - \mathrm{\overline{M}}$, $\mathrm{\overline{M}} - \mathrm{\overline{\Gamma}} - \mathrm{\overline{M}}$, and $\mathrm{\overline{X}} - \mathrm{\overline{\Gamma}} - \mathrm{\overline{X}}$ directions. The grey bands represent the surface projected bulk bands, red bands are surface bands for $\mathrm{Sb}$ square net termination (termination in the experimental data), and the blue bands are the surface states with $\mathrm{NdTe}$ plane termination. The comparison with our experimental data shows that the termination of the sample is on the $\mathrm{GdTe}$ plane. The STEM images for (001) surface of $\mathrm{GdSbTe}$ was presented Ref. \cite{GdSbTeSankarS}, which shows that the $\mathrm{NdTe}$ termination is preferred. It can suggest general features of this type of compounds and termination on the $R\mathrm{Te}$ ($R$ = rare earth) plane.\\

\begin{figure*} [h]
\includegraphics[width=1\textwidth]{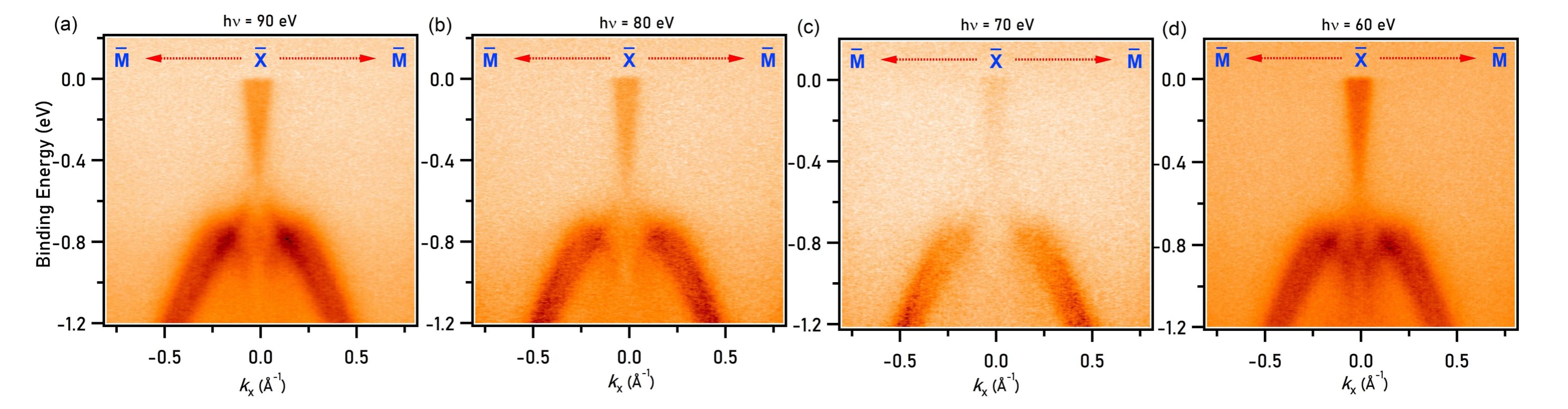}
\caption{ARPES measured band dispersion along $\mathrm{\overline{M}} - \mathrm{\overline{X}} - \mathrm{\overline{M}}$ measured at a photon energy of (a) $\mathrm{90~eV}$, (b) $\mathrm{80~eV}$, (c) $\mathrm{70~eV}$, and (d) $\mathrm{60~eV}$.}
\label{SF6}
\end{figure*}

\begin{center}
\textbf{6. BAND STRUCTURE ALONG $\mathrm{\overline{M}} - \mathrm{\overline{X}} - \mathrm{\overline{M}}$ AT DIFFERENT PHOTON ENERGIES}
\end{center}
In Fig. \ref{SF6}, we present the ARPES measured electronic band dispersion along the $\mathrm{\overline{M}} - \mathrm{\overline{X}} - \mathrm{\overline{M}}$ direction measured at different photon energies as noted on top of each plot. The gapped Dirac-like feature seems to have a little or no change in dispersion with photon energy variation implying the associated bands are coming from the surface. The surface bands co-exist with the bulk bands as seen from the slab calculation in Fig. \ref{SF5} (blue bands).\\

\begin{figure*} [h]
\includegraphics[width=1\textwidth]{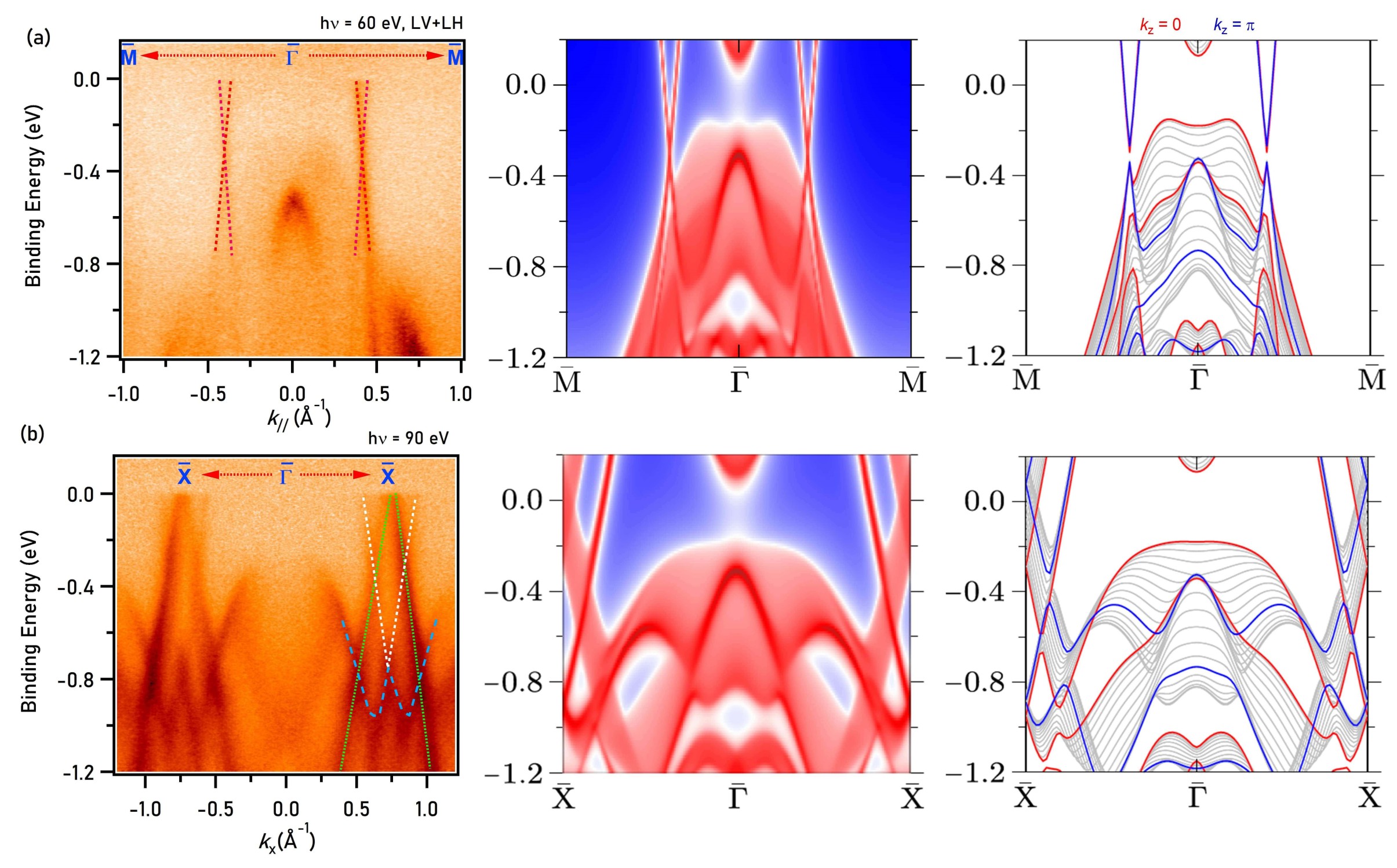}
\caption{ARPES measured band dispersion (left panel), surface projected band structure calculation (middle panel), and $k_z depended$ bulk bands calculation (right panel) along (a) $\mathrm{\overline{M}} - \mathrm{\overline{\Gamma}} - \mathrm{\overline{M}}$ and (b) $\mathrm{\overline{X}} - \mathrm{\overline{\Gamma}} - \mathrm{\overline{X}}$.}
\label{SF7}
\end{figure*}

\begin{center}
\textbf{7. EXPERIMENTAL AND CALCULATED ELECTRONIC STRUCTURE ALONG $\mathrm{\overline{M}} - \mathrm{\overline{\Gamma}} - \mathrm{\overline{M}}$ AND $\mathrm{\overline{X}} - \mathrm{\overline{\Gamma}} - \mathrm{\overline{X}}$}
\end{center}
In Fig. \ref{SF7}, we present the comparison of the experimental and calculated band structure along the $\mathrm{\overline{M}} - \mathrm{\overline{\Gamma}} - \mathrm{\overline{M}}$ and $\mathrm{\overline{X}} - \mathrm{\overline{\Gamma}} - \mathrm{\overline{X}}$ directions. The experimental band dispersion in Fig. \ref{SF7}(a) is obtained from the addition of matrices of LH polarized and LV polarized data. A gapless Dirac crossing can be observed around $\mathrm{300~meV}$ below the Fermi level. This feature is well resolved in calculated band structure. The calculated band structure is the projection from all $k_z$ values and a gap is not resolved, although the bulk band calculations (main text Fig. 1) predict a gap along $\mathrm{\Gamma}-\mathrm{M}$. A small gap can be observed in the bulk bands obtained from slab calculations. The disagreement with the experimental results, where we observe gapless Dirac node, comes from the poor $k_z$ resolution in the vacuum ultra-violet ARPES. The experimental band dispersion along $\mathrm{\overline{X}} - \mathrm{\overline{\Gamma}} - \mathrm{\overline{X}}$ shows a linearly dispersing hole band crossing the Fermi level. This band likely forms the Dirac node above the Fermi level, corresponding to the dispersive nodal-line along the $\mathrm{X}-\mathrm{R}$ direction. Below $\mathrm{800~meV}$ binding energy, another linear crossing exists between an electron band and a W-shaped band. These observations are well reproduced in calculated band calculations. The dispersive nature of the nodal-line in the vicinity of Fermi level and a minimal dispersion of the nodal-line observed below $\mathrm{\sim 800~meV}$ is well supported by the slab calculated bulk band dispersion.

\begin{thebibliography}{50}
\bibitem{TSM} N. P. Armitage, E. J. Mele, and A. Vishwanath, Weyl and Dirac semimetals in three-dimensional solids, \href{https://doi.org/10.1103/RevModPhys.90.015001}{Rev. Mod. Phys. \textbf{90}, 015001 (2018)}.

\bibitem{Na3Bi} Z. K. Liu, B. Zhou, Y. Zhang, Z. J. Wang, H. M. Weng, D. Prabhakaran, S. K. Mo, Z. X. Shen, Z. Fang, X. Dai, Z. Hussain, and Y. L. Chen, Discovery of a Three-Dimensional Topological Dirac Semimetal, $\mathrm{Na}_3\mathrm{Bi}$, \href{https://doi.org/10.1126/science.124508}{Science \textbf{343}, 864 (2014)}.

\bibitem{Cd3As2} M. Neupane, S.-Y. Xu, R. Sankar, N. Alidoust, G. Bian, C. Liu, I. Belopolski, T.-R. Chang, H.-T. Jeng, H. Lin, A. Bansil, F. Chou, and M. Z. Hasan, Observation of a three-dimensional topological Dirac semimetal phase in high-mobility $\mathrm{Cd}_3\mathrm{As}_2$, \href{https://doi.org/10.1038/ncomms4786}{Nat. Commun. \textbf{5}, 3786 (2014)}.


\bibitem{TaAsXu} S.-Y. Xu, I. Belopolski, N. Alidoust, M. Neupane, G. Bian, C. Zhang, R. Sankar, G. Chang, Z. Yuan, C.-C. Lee, S.-M. Huang, H. Zheng, J. Ma, D. S. Sanchez, B. Wang, A. Bansil, F. Chou, P. P. Shibayev, H. Lin, S. Jia, and M. Z. Hasan, Discovery of a Weyl fermion semimetal and topological Fermi arcs, \href{https://doi.org/10.1126/science.aaa929}{Science \textbf{349}, 613 (2015)}.

\bibitem{TaAsDing}	B. Q. Lv, H. M. Weng, B. B. Fu, X. P. Wang, H. Miao, J. Ma, P. Richard, X. C. Huang, L. X. Zhao, G. F. Chen, Z. Fang, X. Dai, T. Qian, and H. Ding, Experimental Discovery of Weyl Semimetal $\mathrm{TaAs}$, \href{https://doi.org/10.1103/PhysRevX.5.031013}{Phys. Rev. X \textbf{5}, 031013 (2015)}.

\bibitem{Weyl} B. Yan and C. Felser, Topological Materials: Weyl Semimetals, \href{https://doi.org/10.1146/annurev-conmatphys-031016-025458}{Annu. Rev. Condens. Matter Phys. \textbf{8}, 337 (2017)}.

\bibitem{WeylII} A. A. Soluyanov, D. Gresch, Z. Wang, Q. Wu, M. Troyer, X. Dai, and B. A. Bernevig, Type-II Weyl semimetals, \href{https://doi.org/10.1038/nature15768}{Nature \textbf{527}, 495 (2015)}.

\bibitem{Nodal} A. A. Burkov, M. D. Hook, and L. Balents, Topological nodal semimetals, \href{https://doi.org/10.1103/PhysRevB.84.235126}{Phys. Rev. B \textbf{84}, 235126 (2011)}.

\bibitem{PbTaSe2} G. Bian, T.-R. Chang, R. Sankar, S.-Y. Xu, H. Zheng, T. Neupert, C.-K. Chiu, S.-M. Huang, G. Chang, I. Belopolski, D. S. Sanchez, M. Neupane, N. Alidoust, C. Liu, B. Wang, C.-C. Lee, H.-T. Jeng, C. Zhang, Z. Yuan, S. Jia, A. Bansil, F. Chou, H. Lin, and M. Z. Hasan, Topological nodal-line fermions in spin-orbit metal $\mathrm{Pb}\mathrm{Ta}\mathrm{Se}_2$, \href{https://doi.org/10.1038/ncomms10556}{Nat. Commun. \textbf{7}, 10556 (2016)}.

\bibitem{ZrSiSNeupane} M. Neupane, I. Belopolski, M. M. Hosen, D. S. Sanchez, R. Sankar, M. Szlawska, S.-Y. Xu, K. Dimitri, N. Dhakal, P. Maldonado, P. M. Oppeneer, D. Kaczorowski, F. Chou, M. Z. Hasan, and T. Durakiewicz, Observation of topological nodal fermion semimetal phase in $\mathrm{ZrSiS}$, \href{https://doi.org/10.1103/PhysRevB.93.201104}{Phys. Rev. B \textbf{93}, 201104 (2016)}.

\bibitem{ZrSiSSchoop} L. M. Schoop, M. N. Ali, C. Straßer, A. Topp, A. Varykhalov, D. Marchenko, V. Duppel, S. S. P. Parkin, B. V. Lotsch, and C. R. Ast, Dirac cone protected by non-symmorphic symmetry and three-dimensional Dirac line node in $\mathrm{ZrSiS}$, \href{https://doi.org/10.1038/ncomms11696}{Nat. Commun. \textbf{7}, 11696 (2016)}.

\bibitem{Bradlyn} B. Bradlyn, J. Cano, Z. Wang, M. G. Vergniory, C. Felser, R. J. Cava, and B. A. Bernevig, Beyond Dirac and Weyl fermions: Unconventional quasiparticles in conventional crystals, \href{https://doi.org/10.1126/science.aaf503}{Science \textbf{353}, aaf5037 (2016)}.

\bibitem{CoSi} D. Takane, Z. Wang, S. Souma, K. Nakayama, T. Nakamura, H. Oinuma, Y. Nakata, H. Iwasawa, C. Cacho, T. Kim, K. Horiba, H. Kumigashira, T. Takahashi, Y. Ando, and T. Sato, Observation of Chiral Fermions with a Large Topological Charge and Associated Fermi-Arc Surface States in $\mathrm{CoSi}$, \href{https://doi.org/10.1103/PhysRevLett.122.076402}{Phys. Rev. Lett. \textbf{122}, 076402 (2019)}.

\bibitem{RhSi} D. S. Sanchez, I. Belopolski, T. A. Cochran, X. Xu, J.-X. Yin, G. Chang, W. Xie, K. Manna, V. Süß, C.-Y. Huang, N. Alidoust, D. Multer, S. S. Zhang, N. Shumiya, X. Wang, G.-Q. Wang, T.-R. Chang, C. Felser, S.-Y. Xu, S. Jia, H. Lin, and M. Z. Hasan, Topological chiral crystals with helicoid-arc quantum states, \href{https://doi.org/10.1038/s41586-019-1037-2}{Nature \textbf{567}, 500 (2019)}.

\bibitem{Co2MnGa} I. Belopolski, K. Manna, D. S. Sanchez, G. Chang, B. Ernst, J. Yin, S. S. Zhang, T. Cochran, N. Shumiya, H. Zheng, B. Singh, G. Bian, D. Multer, M. Litskevich, X. Zhou, S.-M. Huang, B. Wang, T.-R. Chang, S.-Y. Xu, A. Bansil, C. Felser, H. Lin, and M. Z. Hasan, Discovery of topological Weyl fermion lines and drumhead surface states in a room temperature magnet, \href{https://doi.org/10.1126/science.aav23}{Science \textbf{365}, 1278 (2019)}.

\bibitem{Topp} A. Topp, J. M. Lippmann, A. Varykhalov, V. Duppel, B. V. Lotsch, C. R. Ast, and L. M. Schoop, Non-symmorphic band degeneracy at the Fermi level in $\mathrm{ZrSiTe}$, \href{https://doi.org/10.1088/1367-2630/aa4f65}{New J. Phys. \textbf{18}, 125014 (2016)}.

\bibitem{Hu} J. Hu, Z. Tang, J. Liu, X. Liu, Y. Zhu, D. Graf, K. Myhro, S. Tran, C. N. Lau, J. Wei, and Z. Mao, Evidence of Topological Nodal-Line Fermions in $\mathrm{ZrSiSe}$ and $\mathrm{ZrSiTe}$, \href{https://doi.org/10.1103/PhysRevLett.117.016602}{Phys. Rev. Lett. \textbf{117}, 016602 (2016)}.

\bibitem{Takane} D. Takane, Z. Wang, S. Souma, K. Nakayama, C. X. Trang, T. Sato, T. Takahashi, and Y. Ando, Dirac-node arc in the topological line-node semimetal $\mathrm{HfSiS}$, \href{https://doi.org/10.1103/PhysRevB.94.121108}{Phys. Rev. B \textbf{94}, 121108 (2016)}.

\bibitem{ZrSiX} M. M. Hosen, K. Dimitri, I. Belopolski, P. Maldonado, R. Sankar, N. Dhakal, G. Dhakal, T. Cole, P. M. Oppeneer, D. Kaczorowski, F. Chou, M. Z. Hasan, T. Durakiewicz, and M. Neupane, Tunability of the topological nodal-line semimetal phase in $\mathrm{ZrSi}X$-type materials ($X=\mathrm{S}, \mathrm{Se}, \mathrm{Te}$), \href{https://doi.org/10.1103/PhysRevB.95.161101}{Phys. Rev. B \textbf{95}, 161101 (2017)}.

\bibitem{Chen} C. Chen, X. Xu, J. Jiang, S. C. Wu, Y. P. Qi, L. X. Yang, M. X. Wang, Y. Sun, N. B. M. Schröter, H. F. Yang, L. M. Schoop, Y. Y. Lv, J. Zhou, Y. B. Chen, S. H. Yao, M. H. Lu, Y. F. Chen, C. Felser, B. H. Yan, Z. K. Liu, and Y. L. Chen, Dirac line nodes and effect of spin-orbit coupling in the nonsymmorphic critical semimetals $M\mathrm{SiS}\phantom{\rule{0.16em}{0ex}}(M=\mathrm{Hf},\phantom{\rule{0.16em}{0ex}}\mathrm{Zr})$, \href{https://doi.org/10.1103/PhysRevB.95.125126}{Phys. Rev. B \textbf{95}, 125126 (2017)}.

\bibitem{Lou} R. Lou, J. Z. Ma, Q. N. Xu, B. B. Fu, L. Y. Kong, Y. G. Shi, P. Richard, H. M. Weng, Z. Fang, S. S. Sun, Q. Wang, H. C. Lei, T. Qian, H. Ding, and S. C. Wang, Emergence of topological bands on the surface of $\mathrm{ZrSnTe}$ crystal, \href{https://doi.org/10.1103/PhysRevB.93.241104}{Phys. Rev. B \textbf{93}, 241104 (2016)}.

\bibitem{ZrGeTe} M. M. Hosen, K. Dimitri, A. Aperis, P. Maldonado, I. Belopolski, G. Dhakal, F. Kabir, C. Sims, M. Z. Hasan, D. Kaczorowski, T. Durakiewicz, P. M. Oppeneer, and M. Neupane, Observation of gapless Dirac surface states in $\mathrm{ZrGeTe}$, \href{https://doi.org/10.1103/PhysRevB.97.121103}{Phys. Rev. B \textbf{97}, 121103 (2018)}.

\bibitem{Fu} B. B. Fu, C. J. Yi, T. T. Zhang, M. Caputo, J. Z. Ma, X. Gao, B. Q. Lv, L. Y. Kong, Y. B. Huang, P. Richard, M. Shi, V. N. Strocov, C. Fang, H. M. Weng, Y. G. Shi, T. Qian, and H. Ding, Dirac nodal surfaces and nodal lines in $\mathrm{ZrSiS}$, \href{https://doi.org/10.1126/sciadv.aau6459}{Sci. Adv. \textbf{5}, eaau6459 (2019)}.

\bibitem{Ali} M. N. Ali, L. M. Schoop, C. Garg, J. M. Lippmann, E. Lara, B. Lotsch, and S. S. P. Parkin, Butterfly magnetoresistance, quasi-2D Dirac Fermi surface and topological phase transition in $\mathrm{ZrSiS}$, \href{https://doi.org/10.1126/sciadv.1601742}{Sci. Adv. \textbf{2}, e1601742 (2016)}.

\bibitem{Lv} Y.-Y. Lv, B.-B. Zhang, X. Li, S.-H. Yao, Y. B. Chen, J. Zhou, S.-T. Zhang, M.-H. Lu, and Y.-F. Chen, Extremely large and significantly anisotropic magnetoresistance in $\mathrm{ZrSiS}$ single crystals, \href{https://doi.org/10.1063/1.4953772}{Appl. Phys. Lett. \textbf{108}, 244101 (2016)}.

\bibitem{Schilling} M. B. Schilling, L. M. Schoop, B. V. Lotsch, M. Dressel, and A. V. Pronin, Flat Optical Conductivity in $\mathrm{ZrSiS}$ due to Two-Dimensional Dirac Bands, \href{https://doi.org/10.1103/PhysRevLett.119.187401}{Phys. Rev. Lett. \textbf{119}, 187401 (2017)}.

\bibitem{Singha} R. Singha, A. K. Pariari, B. Satpati, and P. Mandal, Large nonsaturating magnetoresistance and signature of nondegenerate Dirac nodes in $\mathrm{ZrSiS}$, \href{https://doi.org/10.1073/pnas.1618004114}{Proc. Natl. Acad. Sci. \textbf{114}, 2468 (2017)}.

\bibitem{Kumar} N. Kumar, K. Manna, Y. Qi, S.-C. Wu, L. Wang, B. Yan, C. Felser, and C. Shekhar, Unusual magnetotransport from Si-square nets in topological semimetal $\mathrm{HfSiS}$, \href{https://doi.org/10.1103/PhysRevB.95.121109}{Phys. Rev. B \textbf{95}, 121109 (2017)}.

\bibitem{Pezzini} S. Pezzini, M. R. van Delft, L. M. Schoop, B. V. Lotsch, A. Carrington, M. I. Katsnelson, N. E. Hussey, and S. Wiedmann, Unconventional mass enhancement around the Dirac nodal loop in $\mathrm{ZrSiS}$, \href{https://doi.org/10.1038/nphys4306}{Nat. Phys. \textbf{14}, 178 (2018)}.

\bibitem{GdSbTeSankar} R. Sankar, I. P. Muthuselvam, K. R. Babu, G. S. Murugan, K. Rajagopal, R. Kumar, T.-C. Wu, C.-Y. Wen, W.-L. Lee, G.-Y. Guo, and F.-C. Chou, Crystal Growth and Magnetic Properties of Topological Nodal-Line Semimetal $\mathrm{GdSbTe}$ with Antiferromagnetic Spin Ordering, \href{https://doi.org/10.1021/acs.inorgchem.9b01698}{Inorg. Chem. \textbf{58}, 11730 (2019)}.

\bibitem{GdSbTe} M. M. Hosen, G. Dhakal, K. Dimitri, P. Maldonado, A. Aperis, F. Kabir, C. Sims, P. Riseborough, P. M. Oppeneer, D. Kaczorowski, T. Durakiewicz, and M. Neupane, Discovery of topological nodal-line fermionic phase in a magnetic material $\mathrm{GdSbTe}$, \href{https://doi.org/10.1038/s41598-018-31296-7}{Sci. Rep. \textbf{8}, 13283 (2018)}.

\bibitem{CeSbTe1}  L. M. Schoop, A. Topp, J. Lippmann, F. Orlandi, L. Müchler, M. G. Vergniory, Y. Sun, A. W. Rost, V. Duppel, M. Krivenkov, S. Sheoran, P. Manuel, A. Varykhalov, B. Yan, R. K. Kremer, C. R. Ast, and B. V. Lotsch, Tunable Weyl and Dirac states in the nonsymmorphic compound $\mathrm{CeSbTe}$, \href{https://doi.org/10.1126/sciadv.aar2317}{Sci. Adv. \textbf{4}, eaar2317 (2018)}.

\bibitem{CeSbTe2} A. Topp, M. G. Vergniory, M. Krivenkov, A. Varykhalov, F. Rodolakis, J. L. McChesney, B. V. Lotsch, C. R. Ast, and L. M. Schoop, The effect of spin-orbit coupling on nonsymmorphic square-net compounds, \href{https://doi.org/10.1016/j.jpcs.2017.12.035}{J.  Phys. Chem.  Solids \textbf{128}, 296 (2019)}.

\bibitem{HoSbTe1} S. Yue, Y. Qian, M. Yang, D. Geng, C. Yi, S. Kumar, K. Shimada, P. Cheng, L. Chen, Z. Wang, H. Weng, Y. Shi, K. Wu, and B. Feng, Topological electronic structure in the antiferromagnet $\mathrm{HoSbTe}$, \href{https://doi.org/10.1103/PhysRevB.102.155109}{Phys. Rev. B \textbf{102}, 155109 (2020)}.

\bibitem{HoSbTe2} M. Yang, Y. Qian, D. Yan, Y. Li, Y. Song, Z. Wang, C. Yi, H. L. Feng, H. Weng, and Y. Shi, Magnetic and electronic properties of a topological nodal line semimetal candidate: $\mathrm{HoSbTe}$, \href{https://doi.org/10.1103/PhysRevMaterials.4.094203}{Phys. Rev. Mater. \textbf{4}, 094203 (2020)}.

\bibitem{DySbTe} F. Gao, J. Huang, W. Ren, M. Li, H. Wang, T. Yang, B. Li, and Z. Zhang, Magnetic and transport properties of the topological compound $\mathrm{DySbTe}$, \href{https://doi.org/10.1103/PhysRevB.105.214434}{Phys. Rev. B \textbf{105}, 214434 (2022)}.

\bibitem{HoSbTe3} N. Shumiya, J.-X. Yin, G. Chang, M. Yang, S. Mardanya, T.-R. Chang, H. Lin, M. S. Hossain, Y.-X. Jiang, T. A. Cochran, Q. Zhang, X. P. Yang, Y. Shi, and M. Z. Hasan, Evidence for electronic signature of a magnetic transition in the topological magnet HoSbTe, \href{https://doi.org/10.1103/PhysRevB.106.035151}{Phys. Rev. B \textbf{106}, 035151 (2022)}.

\bibitem{LaSbTe} Y. Wang, Y. Qian, M. Yang, H. Chen, C. Li, Z. Tan, Y. Cai, W. Zhao, S. Gao, Y. Feng, S. Kumar, E. F. Schwier, L. Zhao, H. Weng, Y. Shi, G. Wang, Y. Song, Y. Huang, K. Shimada, Z. Xu, X. J. Zhou, and G. Liu, Spectroscopic evidence for the realization of a genuine topological nodal-line semimetal in $\mathrm{LaSbTe}$, \href{https://doi.org/10.1103/PhysRevB.103.125131}{Phys. Rev. B \textbf{103}, 125131 (2021)}.

\bibitem{SmSbTe1} S. Regmi, G. Dhakal, F. C. Kabeer, N. Harrison, F. Kabir, A. P. Sakhya, K. Gofryk, D. Kaczorowski, P. M. Oppeneer, and M. Neupane, Observation of multiple nodal lines in $\mathrm{SmSbTe}$, \href{https://doi.org/10.1103/PhysRevMaterials.6.L031201}{Phys. Rev. Mater. \textbf{6}, L031201 (2022)}.

\bibitem{SmSbTe2}  K. Pandey, D. Mondal, J. W. Villanova, J. Roll, R. Basnet, A. Wegner, G. Acharya, M. R. U. Nabi, B. Ghosh, J. Fujii, J. Wang, B. Da, A. Agarwal, I. Vobornik, A. Politano, S. Barraza-Lopez, and J. Hu, Magnetic Topological Semimetal Phase with Electronic Correlation Enhancement in SmSbTe, \href{https://doi.org/10.1002/qute.202100063}{Adv. Quantum Technol. \textbf{4}, 2100063 (2021)}.

\bibitem{NdSbTe1} K. Pandey, R. Basnet, A. Wegner, G. Acharya, M. R. U. Nabi, J. Liu, J. Wang, Y. K. Takahashi, B. Da, and J. Hu, Electronic and magnetic properties of the topological semimetal candidate $\mathrm{NdSbTe}$, \href{https://doi.org/10.1103/PhysRevB.101.235161}{Phys. Rev. B 101, 235161 (2020)}.

\bibitem{CeSbTe3} K. W. Chen, Y. Lai, Y. C. Chiu, S. Steven, T. Besara, D. Graf, T. Siegrist, T. E. Albrecht-Schmitt, L. Balicas, and R. E. Baumbach, Possible devil's staircase in the Kondo lattice $\mathrm{CeSbTe}$, \href{https://doi.org/10.1103/PhysRevB.96.014421}{Phys. Rev. B \textbf{96}, 014421 (2017)}.

\bibitem{CeSbTe4}  B. Lv, J. Chen, L. Qiao, J. Ma, X. Yang, M. Li, M. Wang, Q. Tao, and Z.-A. Xu, Magnetic and transport properties of low-carrier-density Kondo semimetal $\mathrm{CeSbTe}$, \href{https://orcid.org/0000-0001-9290-2762}{J. Phys. Condens. Mater. \textbf{31}, 355601 (2019)}.

\bibitem{CeSbTe5} P. Li, B. Lv, Y. Fang, W. Guo, Z. Wu, Y. Wu, D. Shen, Y. Nie, L. Petaccia, C. Cao, Z.-A. Xu, and Y. Liu, Charge density wave and weak Kondo effect in a Dirac semimetal $\mathrm{CeSbTe}$, \href{https://doi.org/10.1007/s11433-020-1642-2}{Sci. China Phys. Mech. Astron. \textbf{64}, 237412 (2021)}.

\bibitem{NdSbTe2} R. Sankar, I. P. Muthuselvam, K. Rajagopal, K. Ramesh Babu, G. S. Murugan, K. S. Bayikadi, K. Moovendaran, C. Ting Wu, and G.-Y. Guo, Anisotropic Magnetic Properties of Nonsymmorphic Semimetallic Single Crystal $\mathrm{NdSbTe}$, \href{https://doi.org/10.1021/acs.cgd.0c00756}{Crys. Growth  Des. \textbf{20}, 6585 (2020)}.

\bibitem{HK64} P. Hohenber and W. Kohn,  Inhomogeneous Electron Gas,  \href{https://doi.org/10.1103/PhysRev.136.B864}{Phys. Rev. \textbf{136,} B864 (1964)}.

\bibitem{KS65} W. Kohn and L. J. Sham, Self-Consistent Equations Including Exchange and Correlation Effects, \href{https://doi.org/10.1103/PhysRev.140.A1133}{Phys. Rev. \textbf{140}, A1133 (1965)}.

\bibitem{KresseHafner94} G. Kresse and J. Hafner, Ab initio molecular-dynamics simulation of the liquid-metal--amorphous-semiconductor transition in germanium, \href{https://doi.org/10.1103/PhysRevB.49.14251}{Phys. Rev. B \textbf{49}, 14251 (1994)}.

\bibitem{KresseFurthmuller96} G. Kresse and J. Furthmüller, Efficient iterative schemes for ab initio total-energy calculations using a plane-wave basis set, \href{https://doi.org/10.1103/PhysRevB.54.11169}{Phys. Rev. B \textbf{54,} 11169 (1996)}.

\bibitem{KresseJoubert99} G. Kresse and D. Joubert, From ultrasoft pseudopotentials to the projector augmented-wave method, \href{https://doi.org/10.1103/PhysRevB.59.1758}{Phys. Rev. B \textbf{59}, 1758 (1999)}.

\bibitem{Blochl94} P. E. Blöchl, Projector augmented-wave method,  \href{https://doi.org/10.1103/PhysRevB.50.17953}{Phys. Rev. B \textbf{50,} 17953 (1994)}.

\bibitem{SM} For additional information, see the supplemental material to this manuscript, which includes references \cite{NdSbTe1, NdSbTe2, HK64, KS65, KresseHafner94, KresseFurthmuller96, KresseJoubert99, Blochl94, PerdewBurke96, MonkhorstPack76, Pandey2022, Igor2022, LiechtensteinAnisimov95, GiannozziBaroni09, GiannozziAndreussi17, GiannozziBaseggio20, Corso14, MarzhariMostofi12, MarzhariVanderbilt97, SouzaMarzhari01, MostofiYates08, MostofiYates14, PizziVitale20, SanchoSancho85, WuZhang18, GdSbTeSankar}.

\bibitem{Charvillat} J. P. Charvillat, D. Damien, and A. Wojakowski, Crystal chemistry of binary $\mathrm{MSb_2}$ and ternary $\mathrm{MSbTe}$ compounds of transuranium elements, {Revue de Chimie Minerale \textbf{14}, 178 (1977)}.

\bibitem{ZrSiSvf} M. S. Lodge, G. Chang, C.-Y. Huang, B. Singh, J. Hellerstedt, M. T. Edmonds, D. Kaczorowski, M. M. Hosen, M. Neupane, H. Lin, M. S. Fuhrer, B. Weber, and M. Ishigami, Observation of Effective Pseudospin Scattering in $\mathrm{ZrSiS}$, \href{https://doi.org/10.1021/acs.nanolett.7b02307}{Nano Lett. \textbf{17}, 7213 (2017)}.

\bibitem{PerdewBurke96} J. P. Perdew, K. Burke, and M. Ernzerhof, Generalized Gradient Approximation Made Simple, \href{https://doi.org/10.1103/PhysRevLett.77.3865}{Phys. Rev. Lett. \textbf{77}, 3865 (1996)}.

\bibitem{MonkhorstPack76} H. J. Monkhorst and J. D. Pack, Special points for Brillouin-zone integrations, \href{https://doi.org/10.1103/PhysRevB.13.5188}{Phys. Rev. B \textbf{13}, 5188 (1976)}.

\bibitem{Pandey2022} K. Pandey, R. Basnet, J. Wang, B. Da, and J. Hu, Evolution of electronic and magnetic properties in the topological semimetal $\mathrm{Sm}{\mathrm{Sb}}_{x}{\mathrm{Te}}_{2\ensuremath{-}x}$, \href{https://doi.org/10.1103/PhysRevB.105.155139}{Phys. Rev. B \textbf{105}, 155139 (2022)}.

\bibitem{Igor2022} I. Plokhikh, V. Pomjakushin, D. J. Gawryluk, O. Zaharko, and E. Pomjakushina, Competing Magnetic Phases in $Ln\mathrm{SbTe}$ ($Ln$ = $\mathrm{Ho}$ and $\mathrm{Tb}$), \href{https://doi.org/10.1021/acs.inorgchem.2c01711}{Inorg. Chem. \textbf{61}, 11399 (2022)}.

\bibitem{LiechtensteinAnisimov95} A. I. Liechtenstein, V. I. Anisimov, and J. Zaanen, Density-functional theory and strong interactions: Orbital ordering in Mott-Hubbard insulators, \href{https://doi.org/10.1103/PhysRevB.52.R5467}{Phys. Rev. B \textbf{52}, R5467 (1995)}.

\bibitem{GiannozziBaroni09} P. Giannozzi, S. Baroni, N. Bonini, M. Calandra, R. Car, C. Cavazzoni, D. Ceresoli, G. L. Chiarotti, M. Cococcioni, I. Dabo, A. Dal Corso, S. de Gironcoli, S. Fabris, G. Fratesi, R. Gebauer, U. Gerstmann, C. Gougoussis, A. Kokalj, M. Lazzeri, L. Martin-Samos, N. Marzari, F. Mauri, R. Mazzarello, S. Paolini, A. Pasquarello, L. Paulatto, C. Sbraccia, S. Scandolo, G. Sclauzero, A. P. Seitsonen, A. Smogunov, P. Umari, and R. M. Wentzcovitch, QUANTUM ESPRESSO: a modular and open-source software project for quantum simulations of materials, \href{https://doi.org/10.1088/0953-8984/21/39/395502}{J. Phys. Condens. Mater. \textbf{21}, 395502 (2009)}.

\bibitem{GiannozziAndreussi17} P. Giannozzi, O. Andreussi, T. Brumme, O. Bunau, M. Buongiorno Nardelli, M. Calandra, R. Car, C. Cavazzoni, D. Ceresoli, M. Cococcioni, N. Colonna, I. Carnimeo, A. Dal Corso, S. de Gironcoli, P. Delugas, R. A. DiStasio, A. Ferretti, A. Floris, G. Fratesi, G. Fugallo, R. Gebauer, U. Gerstmann, F. Giustino, T. Gorni, J. Jia, M. Kawamura, H. Y. Ko, A. Kokalj, E. Küçükbenli, M. Lazzeri, M. Marsili, N. Marzari, F. Mauri, N. L. Nguyen, H. V. Nguyen, A. Otero-de-la-Roza, L. Paulatto, S. Poncé, D. Rocca, R. Sabatini, B. Santra, M. Schlipf, A. P. Seitsonen, A. Smogunov, I. Timrov, T. Thonhauser, P. Umari, N. Vast, X. Wu, and S. Baroni, Advanced capabilities for materials modelling with Quantum ESPRESSO, \href{https://doi.org/10.1088/1361-648X/aa8f79}{J. Phys. Condens. Mater. \textbf{29}, 465901 (2017)}.

\bibitem{GiannozziBaseggio20} P. Giannozzi, O. Baseggio, P. Bonfà, D. Brunato, R. Car, I. Carnimeo, C. Cavazzoni, S. de Gironcoli, P. Delugas, F. Ferrari Ruffino, A. Ferretti, N. Marzari, I. Timrov, A. Urru, and S. Baroni, Quantum ESPRESSO toward the exascale, \href{https://doi.org/10.1063/5.0005082}{J. Chem. Phys.  \textbf{152}, 154105 (2020)}.

\bibitem{Corso14} A. Dal Corso, Pseudopotentials periodic table: From H to Pu, \href{https://doi.org/10.1016/j.commatsci.2014.07.043}{Comput. Mater. Sci. \textbf{95}, 337 (2014)}.

\bibitem{MarzhariMostofi12} N. Marzari, A. A. Mostofi, J. R. Yates, I. Souza, and D. Vanderbilt, Maximally localized Wannier functions: Theory and applications, \href{https://doi.org/10.1103/RevModPhys.84.1419}{Rev. Mod. Phys. \textbf{84}, 1419 (2012)}.

\bibitem{MarzhariVanderbilt97} N. Marzari and D. Vanderbilt, Maximally localized generalized Wannier functions for composite energy bands, \href{https://doi.org/10.1103/PhysRevB.56.12847}{Phys. Rev. B \textbf{56}, 12847 (1997)}.

\bibitem{SouzaMarzhari01} I. Souza, N. Marzari, and D. Vanderbilt, Maximally localized Wannier functions for entangled energy bands, \href{https://doi.org/10.1103/PhysRevB.65.035109}{Phys. Rev. B \textbf{65}, 035109 (2001)}.

\bibitem{MostofiYates08} A. A. Mostofi, J. R. Yates, Y.-S. Lee, I. Souza, D. Vanderbilt, and N. Marzari, wannier90: A tool for obtaining maximally-localised Wannier functions, \href{https://doi.org/10.1016/j.cpc.2007.11.016}{Comput. Phys. Commun. \textbf{178}, 685 (2008)}.

\bibitem{MostofiYates14} A. A. Mostofi, J. R. Yates, G. Pizzi, Y.-S. Lee, I. Souza, D. Vanderbilt, and N. Marzari, An updated version of wannier90: A tool for obtaining maximally-localised Wannier functions, \href{https://doi.org/10.1016/j.cpc.2014.05.003}{Comput. Phys. Commun. \textbf{185}, 2309 (2014)}.

\bibitem{PizziVitale20} G. Pizzi, V. Vitale, R. Arita, S. Blügel, F. Freimuth, G. Géranton, M. Gibertini, D. Gresch, C. Johnson, T. Koretsune, J. Ibañez-Azpiroz, H. Lee, J.-M. Lihm, D. Marchand, A. Marrazzo, Y. Mokrousov, J. I. Mustafa, Y. Nohara, Y. Nomura, L. Paulatto, S. Poncé, T. Ponweiser, J. Qiao, F. Thöle, S. S. Tsirkin, M. Wierzbowska, N. Marzari, D. Vanderbilt, I. Souza, A. A. Mostofi, and J. R. Yates, Wannier90 as a community code: new features and applications, \href{https://doi.org/10.1088/1361-648X/ab51ff}{J. Phys. Condens. Mater. \textbf{32}, 165902 (2020)}.

\bibitem{SanchoSancho85} M. P. L. Sancho, J. M. L. Sancho, J. M. L. Sancho, and J. Rubio, Highly convergent schemes for the calculation of bulk and surface Green functions,  \href{https://doi.org/10.1088/0305-4608/15/4/009}{J. Phys. F: Met. Phys. \textbf{15}, 851 (1985)}.

\bibitem{WuZhang18} Q. Wu, S. Zhang, H.-F. Song, M. Troyer, and A. A. Soluyanov, WannierTools: An open-source software package for novel topological materials,  \href{https://doi.org/10.1016/j.cpc.2017.09.033}{Comput. Phys. Commun. \textbf{224}, 405 (2018)}.

\end{thebibliography}

\clearpage
\begin{thebibliography}{50}


\bibitem{NdSbTe1S} K. Pandey, R. Basnet, A. Wegner, G. Acharya, M. R. U. Nabi, J. Liu, J. Wang, Y. K. Takahashi, B. Da, and J. Hu, Electronic and magnetic properties of the topological semimetal candidate $\mathrm{NdSbTe}$, \href{https://doi.org/10.1103/PhysRevB.101.235161}{Phys. Rev. B 101, 235161 (2020)}.

\bibitem{NdSbTe2S} R. Sankar, I. P. Muthuselvam, K. Rajagopal, K. Ramesh Babu, G. S. Murugan, K. S. Bayikadi, K. Moovendaran, C. Ting Wu, and G.-Y. Guo, Anisotropic Magnetic Properties of Nonsymmorphic Semimetallic Single Crystal $\mathrm{NdSbTe}$, \href{https://doi.org/10.1021/acs.cgd.0c00756}{Crys. Growth  Des. \textbf{20}, 6585 (2020)}.

\bibitem{HK64S} P. Hohenber and W. Kohn,  Inhomogeneous Electron Gas,  \href{https://doi.org/10.1103/PhysRev.136.B864}{Phys. Rev. \textbf{136,} B864 (1964)}.

\bibitem{KS65S} W. Kohn and L. J. Sham, Self-Consistent Equations Including Exchange and Correlation Effects, \href{https://doi.org/10.1103/PhysRev.140.A1133}{Phys. Rev. \textbf{140}, A1133 (1965)}.

\bibitem{Blochl94S} P. E. Blöchl, Projector augmented-wave method,  \href{https://doi.org/10.1103/PhysRevB.50.17953}{Phys. Rev. B \textbf{50,} 17953 (1994)}.

\bibitem{KresseHafner94S} G. Kresse and J. Hafner, Ab initio molecular-dynamics simulation of the liquid-metal--amorphous-semiconductor transition in germanium, \href{https://doi.org/10.1103/PhysRevB.49.14251}{Phys. Rev. B \textbf{49}, 14251 (1994)}.

\bibitem{KresseFurthmuller96S} G. Kresse and J. Furthmüller, Efficient iterative schemes for ab initio total-energy calculations using a plane-wave basis set, \href{https://doi.org/10.1103/PhysRevB.54.11169}{Phys. Rev. B \textbf{54,} 11169 (1996)}.

\bibitem{KresseJoubert99S} G. Kresse and D. Joubert, From ultrasoft pseudopotentials to the projector augmented-wave method, \href{https://doi.org/10.1103/PhysRevB.59.1758}{Phys. Rev. B \textbf{59}, 1758 (1999)}.


\bibitem{PerdewBurke96S} J. P. Perdew, K. Burke, and M. Ernzerhof, Generalized Gradient Approximation Made Simple, \href{https://doi.org/10.1103/PhysRevLett.77.3865}{Phys. Rev. Lett. \textbf{77}, 3865 (1996)}.

\bibitem{MonkhorstPack76S} H. J. Monkhorst and J. D. Pack, Special points for Brillouin-zone integrations, \href{https://doi.org/10.1103/PhysRevB.13.5188}{Phys. Rev. B \textbf{13}, 5188 (1976)}.

\bibitem{Pandey2022S} K. Pandey, R. Basnet, J. Wang, B. Da, and J. Hu, Evolution of electronic and magnetic properties in the topological semimetal $\mathrm{Sm}{\mathrm{Sb}}_{x}{\mathrm{Te}}_{2\ensuremath{-}x}$, \href{https://doi.org/10.1103/PhysRevB.105.155139}{Phys. Rev. B \textbf{105}, 155139 (2022)}.

\bibitem{Igor2022S} I. Plokhikh, V. Pomjakushin, D. J. Gawryluk, O. Zaharko, and E. Pomjakushina, Competing Magnetic Phases in $Ln\mathrm{SbTe}$ ($Ln$ = $\mathrm{Ho}$ and $\mathrm{Tb}$), \href{https://doi.org/10.1021/acs.inorgchem.2c01711} {Inorg. Chem. \textbf{61}, 11399 (2022)}.

\bibitem{LiechtensteinAnisimov95S} A. I. Liechtenstein, V. I. Anisimov, and J. Zaanen, Density-functional theory and strong interactions: Orbital ordering in Mott-Hubbard insulators, \href{https://doi.org/10.1103/PhysRevB.52.R5467}{Phys. Rev. B \textbf{52}, R5467 (1995)}.

\bibitem{GiannozziBaroni09S} P. Giannozzi, S. Baroni, N. Bonini, M. Calandra, R. Car, C. Cavazzoni, D. Ceresoli, G. L. Chiarotti, M. Cococcioni, I. Dabo, A. Dal Corso, S. de Gironcoli, S. Fabris, G. Fratesi, R. Gebauer, U. Gerstmann, C. Gougoussis, A. Kokalj, M. Lazzeri, L. Martin-Samos, N. Marzari, F. Mauri, R. Mazzarello, S. Paolini, A. Pasquarello, L. Paulatto, C. Sbraccia, S. Scandolo, G. Sclauzero, A. P. Seitsonen, A. Smogunov, P. Umari, and R. M. Wentzcovitch, QUANTUM ESPRESSO: a modular and open-source software project for quantum simulations of materials, \href{https://doi.org/10.1088/0953-8984/21/39/395502}{J. Phys. Condens. Mater. \textbf{21}, 395502 (2009)}.

\bibitem{GiannozziAndreussi17S} P. Giannozzi, O. Andreussi, T. Brumme, O. Bunau, M. Buongiorno Nardelli, M. Calandra, R. Car, C. Cavazzoni, D. Ceresoli, M. Cococcioni, N. Colonna, I. Carnimeo, A. Dal Corso, S. de Gironcoli, P. Delugas, R. A. DiStasio, A. Ferretti, A. Floris, G. Fratesi, G. Fugallo, R. Gebauer, U. Gerstmann, F. Giustino, T. Gorni, J. Jia, M. Kawamura, H. Y. Ko, A. Kokalj, E. Küçükbenli, M. Lazzeri, M. Marsili, N. Marzari, F. Mauri, N. L. Nguyen, H. V. Nguyen, A. Otero-de-la-Roza, L. Paulatto, S. Poncé, D. Rocca, R. Sabatini, B. Santra, M. Schlipf, A. P. Seitsonen, A. Smogunov, I. Timrov, T. Thonhauser, P. Umari, N. Vast, X. Wu, and S. Baroni, Advanced capabilities for materials modelling with Quantum ESPRESSO, \href{https://doi.org/10.1088/1361-648X/aa8f79}{J. Phys. Condens. Mater. \textbf{29}, 465901 (2017)}.

\bibitem{GiannozziBaseggio20S} P. Giannozzi, O. Baseggio, P. Bonfà, D. Brunato, R. Car, I. Carnimeo, C. Cavazzoni, S. de Gironcoli, P. Delugas, F. Ferrari Ruffino, A. Ferretti, N. Marzari, I. Timrov, A. Urru, and S. Baroni, Quantum ESPRESSO toward the exascale, \href{https://doi.org/10.1063/5.0005082}{J. Chem. Phys.  \textbf{152}, 154105 (2020)}.

\bibitem{Corso14S} A. Dal Corso, Pseudopotentials periodic table: From H to Pu, \href{https://doi.org/10.1016/j.commatsci.2014.07.043}{Comput. Mater. Sci. \textbf{95}, 337 (2014)}.

\bibitem{MarzhariMostofi12S} N. Marzari, A. A. Mostofi, J. R. Yates, I. Souza, and D. Vanderbilt, Maximally localized Wannier functions: Theory and applications, \href{https://doi.org/10.1103/RevModPhys.84.1419}{Rev. Mod. Phys. \textbf{84}, 1419 (2012)}.

\bibitem{MarzhariVanderbilt97S} N. Marzari and D. Vanderbilt, Maximally localized generalized Wannier functions for composite energy bands, \href{https://doi.org/10.1103/PhysRevB.56.12847}{Phys. Rev. B \textbf{56}, 12847 (1997)}.

\bibitem{SouzaMarzhari01S} I. Souza, N. Marzari, and D. Vanderbilt, Maximally localized Wannier functions for entangled energy bands, \href{https://doi.org/10.1103/PhysRevB.65.035109}{Phys. Rev. B \textbf{65}, 035109 (2001)}.

\bibitem{MostofiYates08S} A. A. Mostofi, J. R. Yates, Y.-S. Lee, I. Souza, D. Vanderbilt, and N. Marzari, wannier90: A tool for obtaining maximally-localised Wannier functions, \href{https://doi.org/10.1016/j.cpc.2007.11.016}{Comput. Phys. Commun. \textbf{178}, 685 (2008)}.

\bibitem{MostofiYates14S} A. A. Mostofi, J. R. Yates, G. Pizzi, Y.-S. Lee, I. Souza, D. Vanderbilt, and N. Marzari, An updated version of wannier90: A tool for obtaining maximally-localised Wannier functions, \href{https://doi.org/10.1016/j.cpc.2014.05.003}{Comput. Phys. Commun. \textbf{185}, 2309 (2014)}.

\bibitem{PizziVitale20S} G. Pizzi, V. Vitale, R. Arita, S. Blügel, F. Freimuth, G. Géranton, M. Gibertini, D. Gresch, C. Johnson, T. Koretsune, J. Ibañez-Azpiroz, H. Lee, J.-M. Lihm, D. Marchand, A. Marrazzo, Y. Mokrousov, J. I. Mustafa, Y. Nohara, Y. Nomura, L. Paulatto, S. Poncé, T. Ponweiser, J. Qiao, F. Thöle, S. S. Tsirkin, M. Wierzbowska, N. Marzari, D. Vanderbilt, I. Souza, A. A. Mostofi, and J. R. Yates, Wannier90 as a community code: new features and applications, \href{https://doi.org/10.1088/1361-648X/ab51ff}{J. Phys. Condens. Mater. \textbf{32}, 165902 (2020)}.

\bibitem{SanchoSancho85S} M. P. L. Sancho, J. M. L. Sancho, J. M. L. Sancho, and J. Rubio, Highly convergent schemes for the calculation of bulk and surface Green functions,  \href{https://doi.org/10.1088/0305-4608/15/4/009}{J. Phys. F: Met. Phys. \textbf{15}, 851 (1985)}.

\bibitem{WuZhang18S} Q. Wu, S. Zhang, H.-F. Song, M. Troyer, and A. A. Soluyanov, WannierTools: An open-source software package for novel topological materials,  \href{https://doi.org/10.1016/j.cpc.2017.09.033}{Comput. Phys. Commun. \textbf{224}, 405 (2018)}.

\bibitem{GdSbTeSankarS} R. Sankar, I. P. Muthuselvam, K. R. Babu, G. S. Murugan, K. Rajagopal, R. Kumar, T.-C. Wu, C.-Y. Wen, W.-L. Lee, G.-Y. Guo, and F.-C. Chou, Crystal Growth and Magnetic Properties of Topological Nodal-Line Semimetal $\mathrm{GdSbTe}$ with Antiferromagnetic Spin Ordering, \href{https://doi.org/10.1021/acs.inorgchem.9b01698}{Inorg. Chem. \textbf{58}, 11730 (2019)}.


\end{thebibliography}
\end{document}